\documentclass[fleqn,usenatbib]{rasti}

\usepackage{lineno}
\usepackage{comment}

\usepackage{newtxtext,newtxmath}

\usepackage[T1]{fontenc}

\DeclareRobustCommand{\VAN}[3]{#2}
\let\VANthebibliography\thebibliography
\def\thebibliography{\DeclareRobustCommand{\VAN}[3]{##3}\VANthebibliography}

\usepackage{graphicx}	

\title[Ariel and JWST Exoplanet Synergies]{On the synergetic use of Ariel and JWST for exoplanet atmospheric science}

\author[Q. Changeat et al.]{
Q. Changeat$^{1}$\thanks{E-mail: q.changeat@rug.nl},
P-O. Lagage$^{2}$,
G. Tinetti$^{3}$,
B. Edwards$^{4}$,
B. Charnay$^{5}$,
N. B. Cowan$^{6,7}$,
C. Danielski$^{8,9}$,\newauthor
E. Ducrot$^{2,5}$, 
A. Dyrek$^{10,2}$,
M. Ikoma$^{11,12}$,
T. Lueftinger$^{13}$,
G. Micela$^{14}$,
G. Morello$^{15,14}$,
O. Panic$^{16}$, \newauthor
E. Pascale$^{17}$, 
S. Robert$^{18}$,
S. Schleich$^{19}$,
O. Venot$^{20}$,
J. K. Barstow$^{21}$,
A. Bocchieri$^{9,17}$,
J. Y-K. Cho$^{22}$, \newauthor
R. Cloutier$^{23}$, 
A. Coustenis$^{5}$, 
L. Dang$^{24}$, 
Y. Fujii$^{12,25}$,
Y. Ito$^{12,26}$, 
P. Lavvas$^{27,28}$,
Y. Miguel$^{4,29}$, \newauthor
L. V. Mugnai$^{30}$, 
K. H. Yip$^{3}$\thanks{Corresponding author: kai\_hou.yip@kcl.ac.uk}, 
and J. Zak$^{31}$
\\ \\
$^{1}$Kapteyn Institute, University of Groningen, 9747 AD Groningen, NL;
$^{2}$Universit\'e Paris-Saclay, Universit\'e Paris Cit\'e, CEA, CNRS, AIM, F-91191 \\ Gif-sur-Yvette, France;
$^{3}$Department of Physics, King's College London, Strand Building, WC2R 2LS London, UK;
$^{4}$SRON, Space Research Organisation \\ Netherlands, Niels Bohrweg 4, NL-2333 CA, Leiden, The Netherlands;
$^{5}$LIRA, Observatoire de Paris, Universit\'e PSL, CNRS, Sorbonne Université, Univ. \\ Paris Cit\'e, Sorbonne Paris Cité, 92195 Meudon, France;
$^{6}$Department of Physics, McGill University, Montr\'eal QC H3A 2T8, Canada;
$^{7}$Department of \\ Earth \& Planetary Sciences, McGill University, Montr\'eal, QC H3A 0E8, Canada;
$^{8}$Departament d’Astronomia i Astrofisica, University of Valencia,  calle \\ Dr. Moliner, 50, 46100 Burjassot, Valencia, Spain; 
$^{9}$INAF, Osservatorio Astrofisico di Arcetri, 50125 Firenze, Italy;
$^{10}$Space Telescope Science Institute, \\ Baltimore, MD 21218, USA;
$^{11}$ Astrobiology Center, 2-21-1 Osawa, Mitaka, Tokyo 181-8588, Japan;
$^{12}$ National Astronomical Observatory of Japan, \\ 2-21-1 Osawa, Mitaka, Tokyo 181-8588, Japan;
$^{13}$European Space Agency (ESA), European Space Research and Technology Centre (ESTEC), 2201 AZ \\ Noordwijk, NL;
$^{14}$INAF, Osservatorio Astronomico di Palermo, 90134 Palermo, Italy;
$^{15}$ Instituto de Astrofísica de Andalucía (IAA-CSIC), 18008 Granada, \\ Spain
$^{16}$School of Physics and Astronomy, University of Leeds, LS2 9JT, UK;
$^{17}$Dipartimento di Fisica, La Sapienza Università di Roma, I-00185 Roma, \\ Italy;
$^{18}$Royal Belgian Institute for Space Aeronomy (BIRA-IASB), 1180 Uccle, Belgium;
$^{19}$Department of Astrophysics, University of Vienna, \\ Türkenschanzstrasse 17, 1180 Vienna, Austria;
$^{20}$Université Paris Cité and Univ Paris Est Creteil, CNRS, LISA, F-75013 Paris, France;
$^{21}$School of Physical \\ Sciences, The Open University, Milton Keynes, MK7 6AA, UK;
$^{22}$Martin A. Fisher School of Physics, Brandeis University, Waltham, MA 02453, USA; \\
$^{23}$McMaster University, Department of Physics \& Astronomy, Hamilton, ON L8S 4L8, Canada;
$^{24}$Centre for Astrophysics and Department of Physics and \\ Astronomy, University of Waterloo, Waterloo, Ontario, Canada N2L3G1;
$^{25}$Department of Astronomical Science, Graduate University for Advanced Studies \\ (SOKENDAI), Mitaka, Tokyo 181-8588, Japan;
$^{26}$Astrobiology Center, 2-21-1 Osawa, Mitaka, Tokyo 181-8588, Japan;
$^{27}$Laboratoire Environnements et \\ Atmosphères Terrestres et Planétaires, Université Reims Champagne Ardenne, France;
$^{28}$Institut d'Astrophysique de Paris, UMR CNRS 7095, Paris, France; \\
$^{29}$Leiden Observatory, Leiden University, Einsteinweg 55, 2333 CC, Leiden, NL; 
$^{30}$ School of Physics and Astronomy, Cardiff University, Queens Buildings, \\ The Parade, Cardiff CF24 3AA, UK;
$^{31}$Astronomical Institute of the Academy of Sciences, 25165 Ondrejov, Czech Republic
}
\date{Accepted 14 October 2026.}
\pubyear{\the\year{}}

\begin{document}
\label{firstpage}
\pagerange{\pageref{firstpage}--\pageref{lastpage}}

\maketitle

\vspace{-0.5em}
\begin{abstract}
This paper explores the potential for strategic synergies between the James Webb Space Telescope (JWST) and the Ariel Space Telescope, two flagship observatories poised to revolutionise the study of exoplanet atmospheres. Both telescopes have the potential to address common fundamental questions about exoplanets---especially concerning their nature and origins---and serve a growing scientific community. With their operations now anticipated to overlap from 2031, there is a unique opportunity to enhance the scientific outputs of both observatories through coordinated efforts. In this report, authored by the Ariel-JWST Synergy Working Group, part of the Ariel Consortium Science Team, we summarise the capabilities of both telescopes; we highlight their key differences, similarities, synergies, and distinctive strengths. Ariel is designed to conduct a broad survey of exoplanet atmospheres but remains highly flexible, allowing the mission to integrate insights from JWST's discoveries. Findings from JWST, including data from initiatives shaped by NASA's decadal survey priorities and community-driven research themes, will inform the development of Ariel's survey strategy. Conversely, Ariel's broad-wavelength coverage for bright targets provides complementary avenues for exoplanet researchers, particularly those interested in time-domain observations and large-scale atmospheric studies. We identify key pathways for fostering JWST-Ariel synergies, many of which can be initiated even before Ariel's launch. Leveraging their complementary designs and scopes, JWST and Ariel can jointly address fundamental questions about the nature, formation, and evolution of exoplanets. Such collaboration has the potential to maximise the scientific returns of both observatories and  lay the foundation for future facilities  in the roadmap to exoplanet exploration.
\end{abstract}




\section{Introduction}

\begin{figure*}
\centering
    \includegraphics[width = 0.998\textwidth]{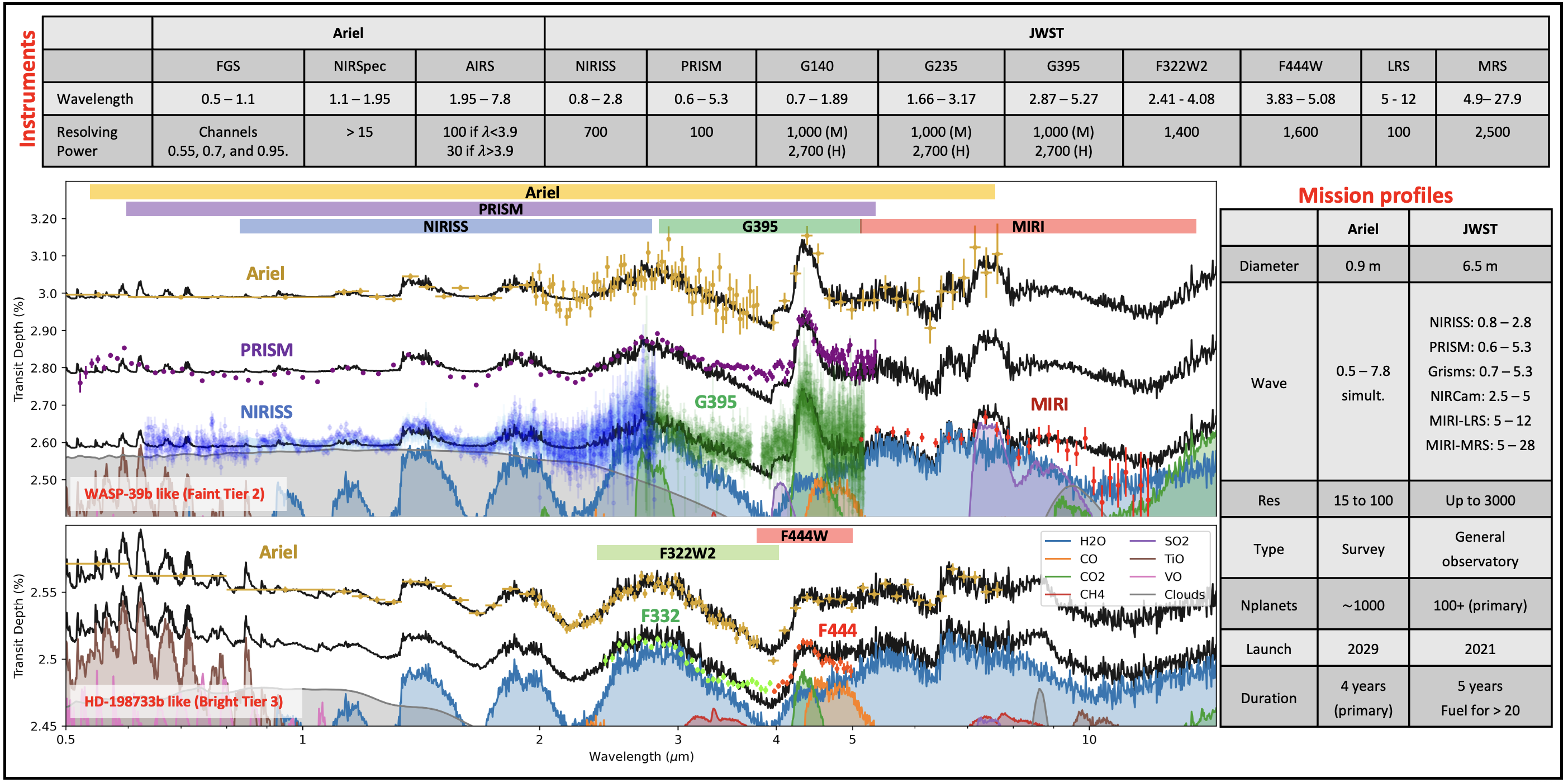}
    \caption{Comparative view of the key performances of the JWST's instruments and Ariel for transiting exoplanets. Top table summarises the characteristics of instruments and observing modes for JWST.  Central left  panel: simulations of a WASP-39\,b-like exoplanet, used as example of a relatively faint target ($V_\mathrm{mag}\sim$ 12) optimal for various JWST's instruments (NIRSpec, NIRISS, MIRI) and  observable by Ariel in Tier 2 (here 2 stacked transits). Bottom left panel: simulation of a HD189733\,b-like exoplanet, used as example of a bright target ($V_\mathrm{mag}\sim$ 7.7) observable with JWST NIRCam  and optimal for Ariel in Tier 3 (here 6 stacked transits). The example data for JWST is from real observations (WASP-39\,b: \citealt{Carter_2024}, \citealt{Powell_2024}; and HD~189733\,b: \citealt{Zhang_2025}) while the displayed Ariel's noise is estimated using the ArielRad simulator \citep{Mugnai_2020}.
   }\label{fig:telescope_characteristics}
\end{figure*}

\begin{figure*}
\centering
    \includegraphics[width = 0.99\textwidth]{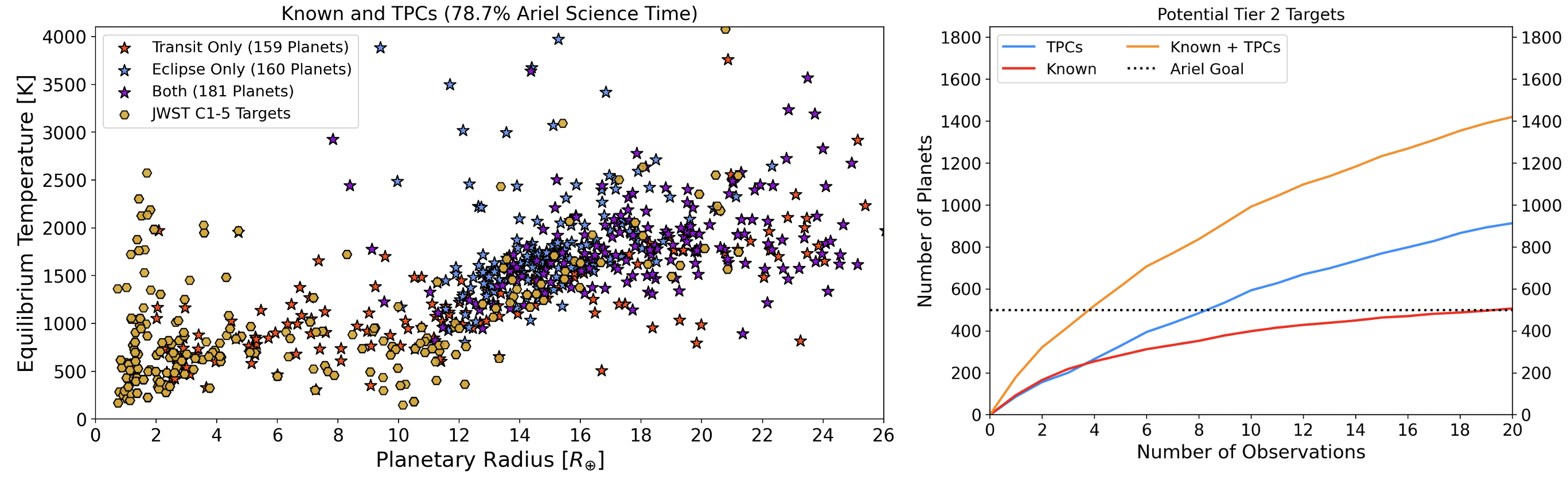}
    \caption{Ariel mission profile. Left: example of a possible Ariel target list for the primary mission (4 years) overlayed over JWST Cycle 1-5 exoplanets; Right: available targets satisfying the Ariel core survey (i.e., Tier 2). Over the primary mission, Ariel should obtain primary transits and secondary eclipses at Tier 2 quality (see example of Tier 2 data quality in Figure \ref{fig:telescope_characteristics}) for about 500 exoplanets---spanning a wide range of planetary radii and temperatures. TPC: TESS Planetary Candidates. Note that TPC can have a high false positive rate; see discussion on predicted Ariel yield in \citet{edwards_ariel2}, from which this figure is inspired.}\label{fig:summary_ariel}
\end{figure*}

Over the past decades, the field of transiting {\it exoplanets} has rapidly grown, transitioning from the detection of a few short orbital period Jupiter-sized exoplanets to the in-depth characterisation of small targets barely larger than the Earth. In 2021, the launch of JWST \citep{gardner_jwst} has marked a turning point for exoplanetary science, providing for the first time high quality data optimised for exoplanet characterisation, i.e. spectra with higher signal-to-noise, higher spectral resolution, higher angular resolution, and broader wavelength coverage than previous facilities. JWST observations nowadays challenge our understanding of exoplanets on a daily basis \citep[][ and reference therein]{Espinoza_2025}. 
In 2031, JWST will be joined at the Lagrange 2 point by the Ariel Space Telescope \citep{Tinetti_ariel, Tinetti_2021_redbook}, an ESA-led mission dedicated to the study of large populations of exoplanet atmospheres. Ariel is a 1\,meter class  telescope, designed with the optimal instrumentation to observe $\sim$1000 atmospheres and use this information to answer fundamental questions about the nature and origins of planetary systems. The main characteristics of Ariel and JWST instruments, as well as example observations, are shown in Figure \ref{fig:telescope_characteristics}. In Figure \ref{fig:summary_ariel}, we also show the targets that are approved for JWST Cycle 1--5 observations and an example target yield for Ariel during its primary mission (four years). The latter predictions are indicative since the final target list will be fixed closer to launch. The predictions are extracted from \cite{edwards_ariel2}, selected with the aim of having diverse transit and eclipse surveys. Ariel's observational strategy and target candidates are expected to evolve until launch and beyond, and  will therefore account for JWST's discoveries and the evolving interests of the community. Given the complementary capabilities of Ariel and JWST for studying exoplanet atmospheres, the Ariel Mission Consortium has established a working group with the aim to study the potential synergies between those two observatories.

The present paper summarises a number of opportunities  that have been identified by this working group for a synergistic use of JWST and Ariel. This initiative was pursued in the context of the Ariel's ``dry-run'' activities\footnote{A trial rehearsal of the Ariel mission, conducted primarily through simulations, is currently underway to ensure the consortium’s readiness for mission launch.}, in parallel to other programs and strategies focused on exoplanets with JWST (see e.g., NASA decadal survey and STScI report on the strategic use of JWST and HST). Section 2 provides an overview of Ariel and its observing strategy, as well as current trends in exoplanet atmospheres with JWST. Section 3 focuses on the technical differences between JWST and Ariel and describes avenues for synergies.

\section{Context and missions' characteristics}

\subsection{Overview of JWST and Ariel}

JWST and Ariel were not designed with the same objectives in mind. JWST is a flagship, multi-purpose observatory serving all the fields of astronomy. Exoplanet-related science is one of the main themes for JWST, with about 20--30\% of the telescope time so far being used for this science \citep{Espinoza_2025}: this is the second largest  theme after galaxies. With its large collecting area  (6.5\,m telescope), JWST provides unparalleled sensitivity to conduct in-depth characterisation of exoplanetary atmospheres. 

Ariel is the first space telescope optimised and completely dedicated to exoplanet atmospheric science: it will function as a survey mission, covering a large population of  exoplanets (500--1000 exoplanets in 4--6 year lifetime) ranging from super-Earths to large gas giants, in various irradiation conditions.   Such a large survey will enable us to build a robust statistical understanding of the exoplanet population  \citep{edwards_ariel,edwards_ariel2}.  

Figure \ref{fig:telescope_characteristics}  summarises the instruments' characteristics, missions' profiles, and capabilities of Ariel and JWST. Ariel capabilities are simulated using the radiometric model ArielRad \citep{Mugnai_2020}. The tables are accompanied with simulated spectra for two hot-Jupiters, selected for 
the different brightness of their host stars, i.e. WASP-39 ($V_\mathrm{mag}\sim$ 12) and HD-189733 ($V_\mathrm{mag}\sim$ 7.7). The figure shows overlapping capabilities as well as key differences. Both observatories provide significantly higher quality datasets compared to previous generation of space infrared instruments used to observe exoplanetary atmospheres (HST-WFC3, Spitzer). Thanks to a near perfect launch by Arianespace, JWST will have the opportunity to operate for much longer than expected,  well into the 2030s, therefore allowing to expand the exoplanet science program, e.g. by increasing the number of targets observed compared to predictions, and/or through deeper characterisation of individual objects. Most importantly, the foreseen temporal overlap with Ariel will enable new avenues  for synergistic programs.

\subsection{Ariel's observational strategy}\label{sec:ariel_strat}
Ariel will use three photometers---VISPhot (0.5–0.6 $\mu$m), FGS1 (0.6–0.8 $\mu$m), and FGS2 (0.8–1.0 $\mu$m)---and two spectrographs: NIRSpec (1.0–1.95 $\mu$m) and AIRS, with two channels covering 1.95–3.9 $\mu$m and 3.9–7.8 $\mu$m \cite[see:][]{Tinetti_2021_redbook}. These will be used simultaneously to cover the 0.5--7.8\,$\mu$m region in one shot. Such a wide wavelength coverage is required to build a comprehensive view of exo-atmospheres and probe a wide range of atmospheric processes, as well as to properly correct for the stellar signal and instrumental systematics. In Figure \ref{fig:telescope_characteristics}, we show the relevance of the chosen wavelength coverage for Ariel to constrain e.g., clouds, key molecules (e.g., H$_2$O, CO, CO$_2$, CH$_4$), and photochemical products (such as SO$_2$). While the larger collecting area of JWST allows higher signal-to-noise (SNR) in shorter time to be obtained, the duration of a transit/eclipse observation cannot be shortened. Ideally, the need for out-of-transit or out-of-eclipse baseline requires the total observing time to be twice the event duration at minimum. Given that multiple JWST's observations are required to cover the wavelength range of Ariel, Ariel can be more efficient than JWST to characterise exoplanets orbiting very bright stars. This is not the case for fainter targets, where multiple visits are required to compensate for its smaller 1-meter class telescope compared to JWST.

Regarding the selection of targets, Ariel will prioritise the diversity of the sample. This implies that some observationally challenging targets---but exotic, or rare, or required to diversify the surveyed population---will also have to be observed, potentially multiple times to increase their spectral quality. Ariel's observing strategy is based on a tier system: we summarise here the main points for convenience but refer to e.g., \cite{Tinetti_2021_redbook, Changeat_2020_alfnoor, Mugnai_2021_alfnoor, edwards_ariel2, Bocchieri_2023, Cowan_2025} for more details and examples of Ariel surveys. Tier 1 is a {\it reconnaissance} survey. These observations aim to obtain a quick preview  of the full mission sample, offering the opportunity to refine the planetary, stellar and orbital parameters and to identify the best targets for further atmospheric follow-up in other Tiers. In Tier 1, targets must reach an average SNR $> 7$ over five spectral elements (i.e., low resolution criteria)\footnote{The SNR used for Ariel is defined in \cite{Mugnai_2020}. In transit, the contrast ratio (or signal S) is five scale heights; In eclipse, the contrast ratio is the sum of planetary emission and reflection over the stellar flux.}. Over the first four years (primary mission), Ariel aims to collect data for $\sim 1000$ targets in Tier 1. Tier 2 represents the {\it core survey}. Tier 1 targets are here re-observed to build the required SNR for atmospheric studies and conduct the main population survey. In Tier 2, the spectra should reach SNR $> 7$ at resolutions 10, 50, and 15 in respectively NIRSpec, AIRS1, and AIRS2. About 500 targets  will be observed in Tier 2 during the primary mission. Tier 3 includes \emph{benchmark targets}. For these targets, high SNR is achievable in one or two visits and at full spectral resolution, making them ideal laboratories to test atmospheric circulation models and variability through time. Tier 3 targets will be observed repeatedly to build an in-depth picture of their atmosphere spatially and temporally. About 50--100 objects are expected to be observed in Tier 3. Given the increasing interest in the Ariel team to consider more phase-curve observations---as opposed to only the transit or eclipse events---during the primary mission, some of the Tier priorities might be revised before launch. Note that Ariel  will obtain the same simultaneous 0.5--7.8 $\mu$m coverage independently from the tier and the number of repeated observations. Target selection for the Ariel Tier 1, 2, and 3 surveys will be coordinated by the Ariel Consortium based on predefined criteria---including e.g., observability, expected signal-to-noise, host star and planetary properties, and sample diversity---as well as the evolving pool of known exoplanets to maximize the science objectives of the mission \citep{Tinetti_2021_redbook}. Periodic updates as new discoveries become available are also planned.

\subsection{Highlights from the first five years of JWST's operations}

\begin{figure*}
\centering
    \includegraphics[width = 0.99\textwidth]{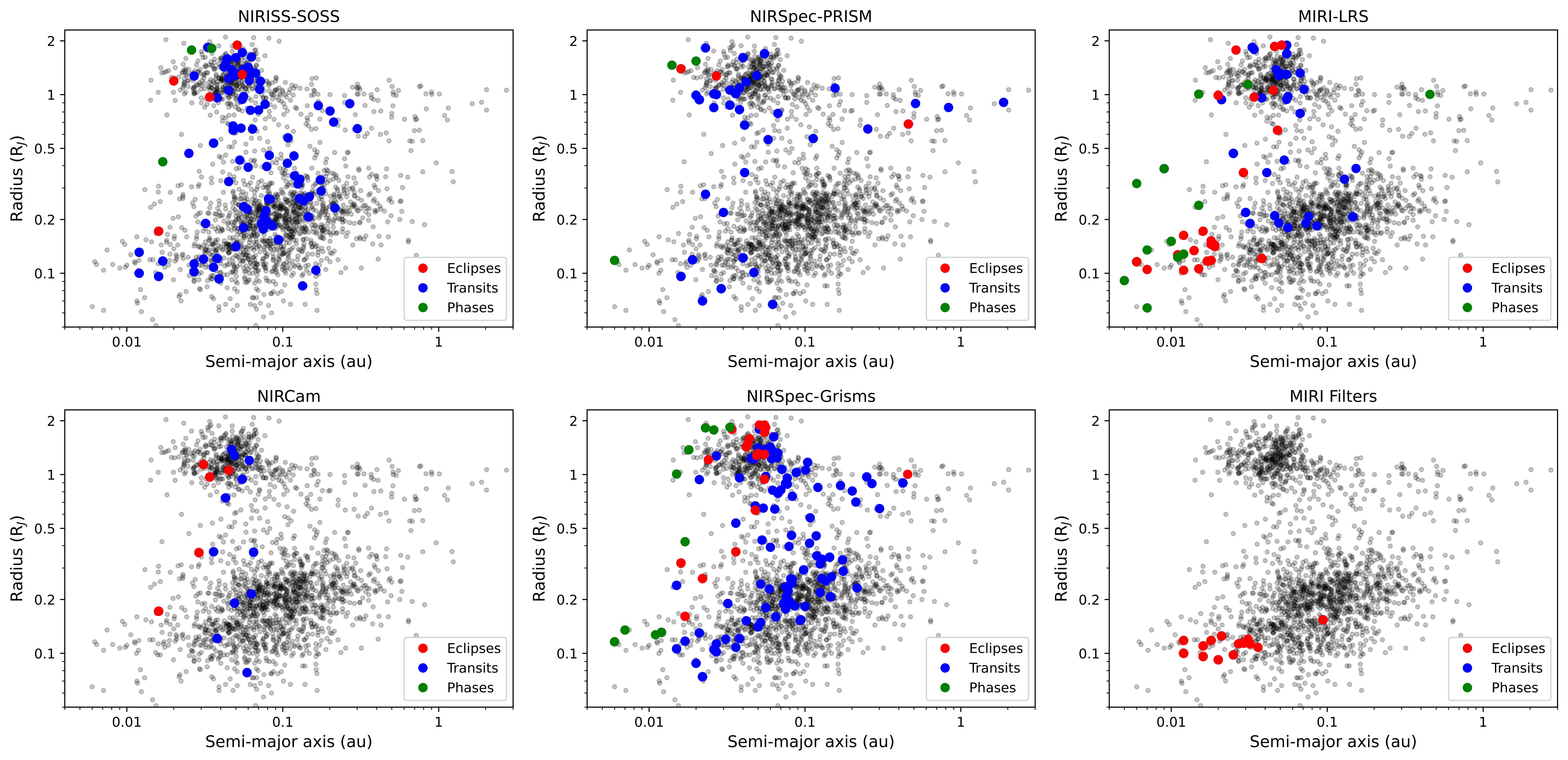}
    \caption{Summary of JWST observations in cycles 1 to 5. The different panels indicate the instruments used (NIRISS, NIRSpec, NIRCam, or MIRI) with background dots showing all currently known exoplanets. Note: NIRCam includes F322W2, and F444W; NIRSpec-Grisms include G235M, G235H,  G395M, and G395H; MIRI filters include FW1280W, and FW1500W. Not all exoplanet observations are show here, for instance HD-189733\,b and 55\,Cnc\,e have been observed via MIRI-MRS}\label{fig:combined_jwst}
\end{figure*}

Since its launch in 2021, JWST has already collected 100s of observations for exoplanetary science. Figure \ref{fig:combined_jwst} provides a view of the transit, eclipse, and phase-curve observations that have been approved for JWST's observations during the first five cycles (note that direct imaging is not shown in the figure). Across these cycles, accepted observations highlight the community's interest for a wide range of targets, sampling a broad range of planetary parameters. Through Cycle 5, JWST will have observed about 220 unique targets: 171 in transit, 64 in eclipse, and 27 in phase-curve (some targets are observed using multiple methods and instruments). Figure \ref{fig:combined_jwst} also illustrates the exoplanet community's preferences amoung the JWST instruments. For example, NIRISS and NIRSpec are currently the most utilized instruments for transiting exoplanets, being mostly employed for transit observations and being less used in eclipse or phase-curve studies. In contrast, MIRI is often utilised for eclipse and phase-curve observations, with also a stronger focus on small objects. These preferences reflect the distinct capabilities of each instrument and the specific science cases prioritised by the exoplanet community so far.

Given the high quality of  JWST's observations and the growth of the exoplanet field over the last decade, a working group was recently formed by STScI on {\it Strategic Exoplanet Initiatives with HST and JWST}. This group aims to sound the community's feedback and to evaluate the strategic use of JWST and HST for exoplanetary science. The working group was also tasked to identify key scientific themes for the potential allocation of 500\,h to exoplanetary science, as Director Discretionary Time (DDT) program. Based on the white papers  received  from the community, this working group identified three priority areas. These were summarised in a recent report \citep{Redfield_2024} and are briefly described here for convenience: \\

\begin{enumerate}

\item Understanding the prevalence and diversity of atmospheres on rocky-M dwarf worlds.

\item Unveiling and understanding population-level trends in gaseous exoplanets. 

\item Putting exoplanets in the context of their stellar environments.

\end{enumerate}

\noindent While the working group found that significant interest was expressed by the exoplanet community for all three science themes, they highlighted science theme 1) as the best option for a 500\,h DDT program with JWST due to its high scientific potential, timely nature, and the difficulty of executing such a challenging program using the uncoordinated General Observer calls. The report recommended to focus the DDT  on an eclipse  survey of rocky worlds orbiting M-dwarfs,  using the MIRI 15\,$\mu$m filter. This survey will select preferably temperate to warm objects---as hotter rocky exoplanets receiving higher radiation could have more tenuous atmospheres---and would allow to identify detectable atmospheres among the chosen sample. The report also recommended  complementary observations with HST (about 240 orbits) to understand the stellar environment.

\begin{figure}
\centering
    \includegraphics[width = 0.45\textwidth]{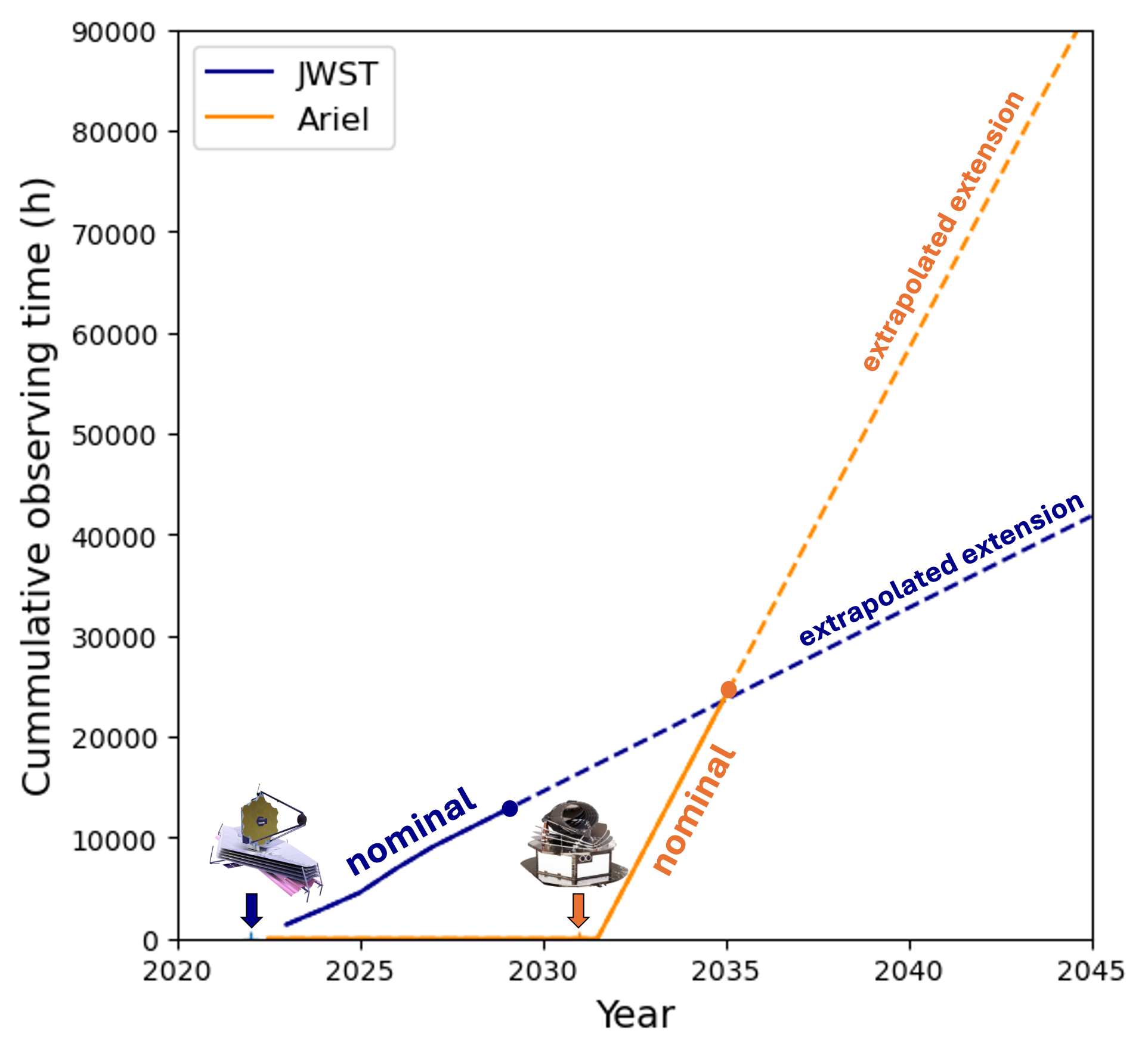}
    \caption{Observing time projections (solid line: nominal mission; dashed line: extended mission) for Ariel and JWST on exoplanet science over the 2030s. We assumed an average of $\sim 6800$\,h per year for Ariel corresponding to the predicted $\sim24000$\,h over 3.5 year of data collection \citep{Morales_2022}. For Ariel, the extended mission is currently 6 years (0.5 year commissioning + 3.5 year nominal + 2 year extended). The data for JWST is obtain from the STScI-JSTUC TAC summaries \citep{JWSTApprovedPrograms}. For JWST, we use the time spent on exoplanets only and extended the projections using the average from Cycle 1 to 5, which is 1818\,h per year.  }\label{fig:ariel_projection}
\end{figure}

The report also highlighted the  interest of the community to pursue large scale surveys of exoplanetary atmospheres, suggesting a 10$^4$\,h  exoplanet survey with JWST. At the current rate of exoplanet observations, JWST is anticipated to acquire $\sim 30,000$\,h of exoplanet data by cycle 20 (assuming the mission receives 16 yearly calls similar to the recent JWST Cycle 5), which would provide statistically significant datasets to address fundamental science questions across the exoplanet population. Note that the latter 30,000\,h is comparable to the total available science time in the Ariel's 4 year nominal mission ($\sim$25,000\,h) as shown in Figure \ref{fig:ariel_projection}. \\

JWST operates as a competitive, proposal-driven observatory. As a result, approved exoplanet observations often emphasize unique targets---such as the smallest, coldest, or highest signal-to-noise ratio (SNR) planets---over more typical ones. In contrast, Ariel is designed as a survey mission, intentionally focusing on representative exoplanet targets. This fundamental difference in strategy leads to naturally complementary observational datasets between the two missions.

\section{Synergies}
In this section we discuss potential key ideas to develop synergies between JWST and Ariel.

\subsection{Science questions common to Ariel and JWST}

JWST's mission is shaped around four main themes: the early universe, the assembly of galaxies, the birth of stars and planetary systems, planetary systems and the origins of life. The last theme covers both extra-solar planets and solar system science, but many questions are common to Ariel's and JWST's objectives. Below is a non-exhaustive list of common sub-themes and science questions pertaining to the two observatories: 
\begin{itemize}
\item Planetary formation. How do planets and planetary systems form and evolve? 
\item
 Physics of atmospheres. What are the physical and chemical processes shaping planetary atmospheres? 
\item
 Composition of exoplanets. What are exoplanets made of? 
\item
 Links with the solar system. Can exoplanets provide clues about the origins of our solar system? 
\item
 Conditions for habitability. What can we understand about planet formation, evolution, and the suitability of planets as habitats? 
\end{itemize}

\begin{figure*}
\centering
    \includegraphics[width = 0.99\textwidth]{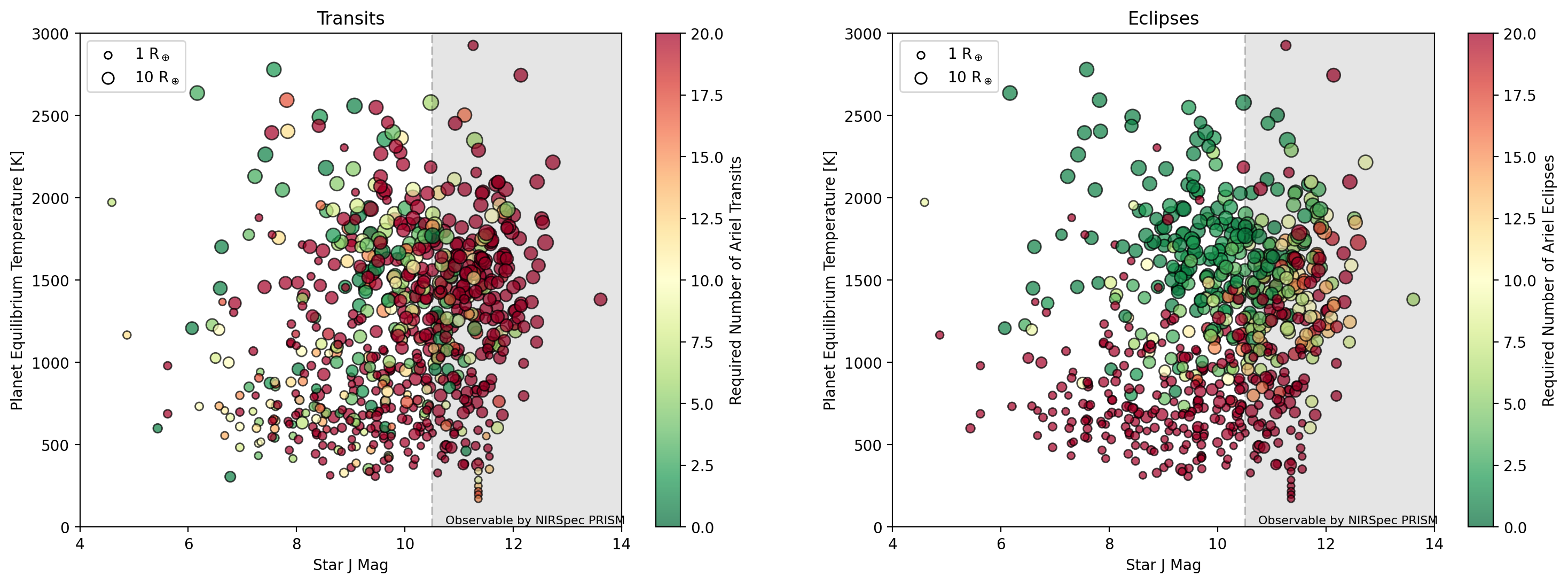}
    \caption{ Ariel target candidates plotted  as a function of planetary temperature and stellar brightness (J mag). The size of the datapoints is proportional to the planetary radius. Colours indicate the required number of transit or eclipse events per target to reach Ariel's Tier 2 data quality. Targets in the shaded area are observable with NIRSpec-PRISM. For transit observations, brightness and planetary size are the key discriminant between Ariel and JWST; for eclipse observations, 
it is rather brightness and planetary temperature. }\label{fig:population_ariel}
\end{figure*}

\subsection{Learning from JWST} \label{sec:learning}

Ariel will greatly benefit from the experience and expertise gained from JWST and other facilities. JWST is currently making important discoveries that may  influence directly the observing strategy of Ariel. We list here a few examples: 
\begin{enumerate}
\item Development of better performing tools and models. With the flow of high-quality data coming from JWST, a significant effort is being made  to upgrade data reduction methods, atmospheric models and retrievals, and associated numerical tools \cite[see e.g., ][]{Bell_2022_eureka, Bushouse_2023, Constantinou_2023, Holmberg_2023, Sarkar_2024, Welbanks_2024, Espinoza_2024, Schleich_2025, Changeat_2025, Challener_2025, Heinke_2026}. 
Ariel will benefit from this acquired knowledge and modelling infrastructure \citep{Mugnai_2025, Mugnai_2026}.
\item
Identification of atmospheric trends from JWST's observations. Preliminary identification of atmospheric trends  will provide valuable information for Ariel. While the nature of JWST's proposal cycles makes it difficult to design large population studies covering all the parameters of interest, trends have started to emerge \cite[see e.g., ][and references therein]{Kempton_2024, Chen_2024, Espinoza_2025, Fu_2025}. For instance, many JWST's observations of hot exoplanets show prominent H$_2$O and CO$_2$ spectral features, indications of photochemical processes, as well as Si-based clouds \citep[e.g., ][]{2023Natur.614..649J, Grant_2023, Dyrek_2023, Fu_2025}. However, a large population (i.e., 50+ objects) covering a wide range of stellar and planetary parameters is likely necessary to fully understand all the driving mechanisms: a science case perfectly fitting Ariel's mission design. The identification of possible chemical trends with JWST,  would allow to tailor Ariel's observations to confirm or refute them on a larger sample of planets \citep{Tinetti_2021_redbook, Venot2025ChemWG, Ito2026SmallPlanets}.   \\

\item
Stellar variability and characterisation. The challenges encountered in understanding and removing  the effects of stellar contamination from transit spectra of small planets---especially those orbiting active late M-dwarfs---will inform the selection of Ariel's targets \citep{Thompson_2026_ariel_sensitivity}. The observing and correction strategies learned from JWST can also be adapted for Ariel \cite[see e.g., Trappist-1 case:][]{Howard_2023, Lim_2023, Allen_2024jwst.prop, Trappist_2024}. \\

\item
Atmospheric temporal and spatial variability. Exoplanets are expected  to show atmospheric temporal variability  \citep{Cho_2003} and three dimensional effects \citep{Caldas_2019, Espinoza_2024, Valentine_2025}. However, we do not know what is the level and timescale of variability in exoplanets, and what is the best strategy to observe these effects through time.
While Ariel will dedicate phase-curve and Tier 3 observations to address these questions \citep{Charnay_2022, Edwards_2026_phasecurve}, any preview from JWST would allow refinement of Ariel's current strategy to monitor more effectively these effects. \\

\item
Community involvement and coordination. Beyond the scientific lessons discussed above, the experience gained in community-led JWST programs can also be useful to Ariel. For instance, the Rocky Worlds DDT program \citep{Redfield_2024} and the JWST Early Release Science (ERS) program \citep{Bean_2018} also illustrate how coordinated observations can influence community practices, including the development of analysis methods, benchmark datasets, and shared software. Although Ariel has a different mission concept and observing strategy, consideration of similar mechanisms for early community coordination and validation are essential for making Ariel data exploitation successful. \\

\item Ephemeris and system parameter maintenance. Both missions depend critically on accurate transit ephemerides and system parameters. Ephemeris degradation is a known operational cost for JWST, and mitigating it has required sustained effort from the community, including ground-based monitoring campaigns \citep{Zellem_2020, Kokori_2022}.
Ariel will inherit both the refreshed ephemerides produced during the first decade of JWST operations and the community practices developed to maintain them. Conversely, effort on the Ariel side to prepare the survey will provide a homogeneous update of the
ephemerides and system parameters of a large fraction of the transiting population, which will directly benefit JWST programs.
\end{enumerate}

\subsection{Ariel and JWST/NIRSpec-PRISM: complementary coverage across host star brightness}\label{sec:bright}

Thanks to its optimal instrumentation and the fact that transit duration cannot be shortened, Ariel can efficiently obtain a broad wavelength coverage (0.5--7.8 $\mu$m) spectra for bright sources. Such a broad spectral coverage is very helpful to capture complex atmospheric processes, probe the vertical structure, and mitigate the effects of stellar contamination \citep{Tinetti_2021_redbook}. Figure \ref{fig:population_ariel} shows a map of optimal targets for Ariel. We also highlight the broad regions in terms of stellar magnitude where JWST provides a significant advantage. For fainter targets, Ariel's performance drops due to its smaller aperture, but a wide wavelength coverage can be obtained with JWST/NIRSpec-PRISM (0.6--5.3 $\mu$m).
This is shown for instance in Figure \ref{fig:telescope_characteristics}, where a WASP-39\,b-like planet (i.e., example of faint target) and a HD-189733\,b-like planet (i.e., example of bright target) are simulated. Many bright targets (mag J $<$ 7) are well suited for Ariel, while broad wavelength range observations can be challenging for JWST due to detector saturation, although newer observing modes---including super-stripe, multi-stripe, and DHS modes---extend JWST's capabilities towards brighter stars \cite[see e.g., ][]{Dyrek_2026}. Exoplanets like HD-189733\,b, HD-209458\,b, GJ-1214\,b, and 55-Cnc\,e are prime Ariel targets (they reach the Tier 2 criteria with a single observation) due to their bright host stars. However, as previously mentioned, Ariel's sensitivity falls rapidly with brightness: planets around stars at mag J $>$ 11 almost always require more than 10 transits to reach Ariel's Tier 2 quality spectra. Such target can (and should) still be observed if it provides important diversity to the Ariel sample, but often, a single PRISM observation can be much more informative. The large wavelength coverage of Ariel (0.5--7.8 $\mu$m) is therefore advantageous for snapshots of the bright targets and there is strong complementarity with JWST, often able to reach the faintest targets, therefore increasing target options in a combined JWST+Ariel population. 

\subsection{JWST: a general observatory for challenging observations}

\subsubsection*{Time consuming targets for Ariel}\label{sec:large_collect}


Due to its high sensitivity, JWST offers significant advantage compared to any other telescope to observe challenging planets that are faint, cold, small, or high gravity. The targets that are most favourable to JWST's observations, but which are costly for Ariel, can be broadly separated in three categories: 
\begin{enumerate}
\item Temperate rocky exoplanets: while significant effort is undertaken to observe the atmosphere of small temperate exoplanets, their study is generally limited by the SNR. In transit, temperate rocky exoplanets are particularly difficult due to their compact atmospheres: hot Jupiters have scale heights around $10-100\times$ larger than temperate super-Earths. In eclipse, their cold temperature lead to $\sim$10\,ppm level signals and spectroscopically resolving molecular features pushes the boundaries of current instruments. For these targets, detecting the signature of an atmosphere is particularly challenging and research teams often adopt fine-tuned observational setups and strategies to maximize the SNR for a particular target. Targeted  observations  for the prominent CO$_2$ feature at 4.5$\mu$m in transit or to constrain the energy redistribution at longer wavelengths  from repeated eclipse measurements were recommended strategies in the 500h DDT survey \citep{Redfield_2024}. While some small exoplanets are good targets for Ariel (see also observing strategies in Section \ref{sec:rockies}), JWST is uniquely positioned to pursue observations of small temperate planets as shown by the effort made for $< 2$\,R$_{\oplus}$ targets in cycles 1--5 (see Figure \ref{fig:combined_jwst}).
\item Planets orbiting faint stars: stellar hosts  with magnitudes J $>$ 11 are challenging  for Ariel, requiring stacking of multiple observations to boost the 
atmospheric signals  of their planetary companions (see Section \ref{sec:bright}). However, these can be prime targets for the JWST/NIRSpec-PRISM instrument \citep{Jakobsen_2022_ns, Birkmann_2022, Boker_2023}. 
\item 
Eclipses: many targets can be observed by Ariel in eclipse, but the atmospheric eclipse signal can be smaller than in transit, especially for colder planets. 
Thanks to its higher sensitivity, JWST can observe  more easily colder targets in eclipse. In addition, JWST/NIRISS \citep{Albert_2023, Doyon_2023} eclipse observations in the optical/NIR can be used to record reflected spectra. 
\end{enumerate}

\subsubsection*{Complementary population: observations of giant planets at large orbital separation}

JWST is a general observatory capable of performing high-contrast imaging observations of exoplanets \citep{Hinkley_2022, Carter_2023_di, Miles_2023}. This population of young
giant planets and brown dwarfs is not accessible to Ariel (i.e., direct imaging is not part of the Ariel core survey), but their study is important to build a comprehensive view of planet formation and atmospheric physics. A systematic comparison between directly imaged young gas giants and short-period transiting planets will shed light on the dynamical and chemical consequences of planetary migration. In particular, accounting for the inclusion of stellar effects, e.g, the role of extreme irradiation when the planets reach close-in orbits, is fundamental to assessing how migration modifies atmospheric composition. This approach can help placing tighter constraints on theoretical models of planet formation, migration, and subsequent evolutionary pathways.
Additionally, due to the particularly high signal-to-noise that can be obtained from directly imaged spectra (see e.g., VHS 1256-1257 b in Section \ref{sec:highres}), these planets are  the perfect test-beds for  exo-atmospheric models.

\subsection{Super-Earths and sub-Neptune populations with Ariel}\label{sec:rockies}

\begin{figure*}
\centering
    \includegraphics[width = 0.99\textwidth]{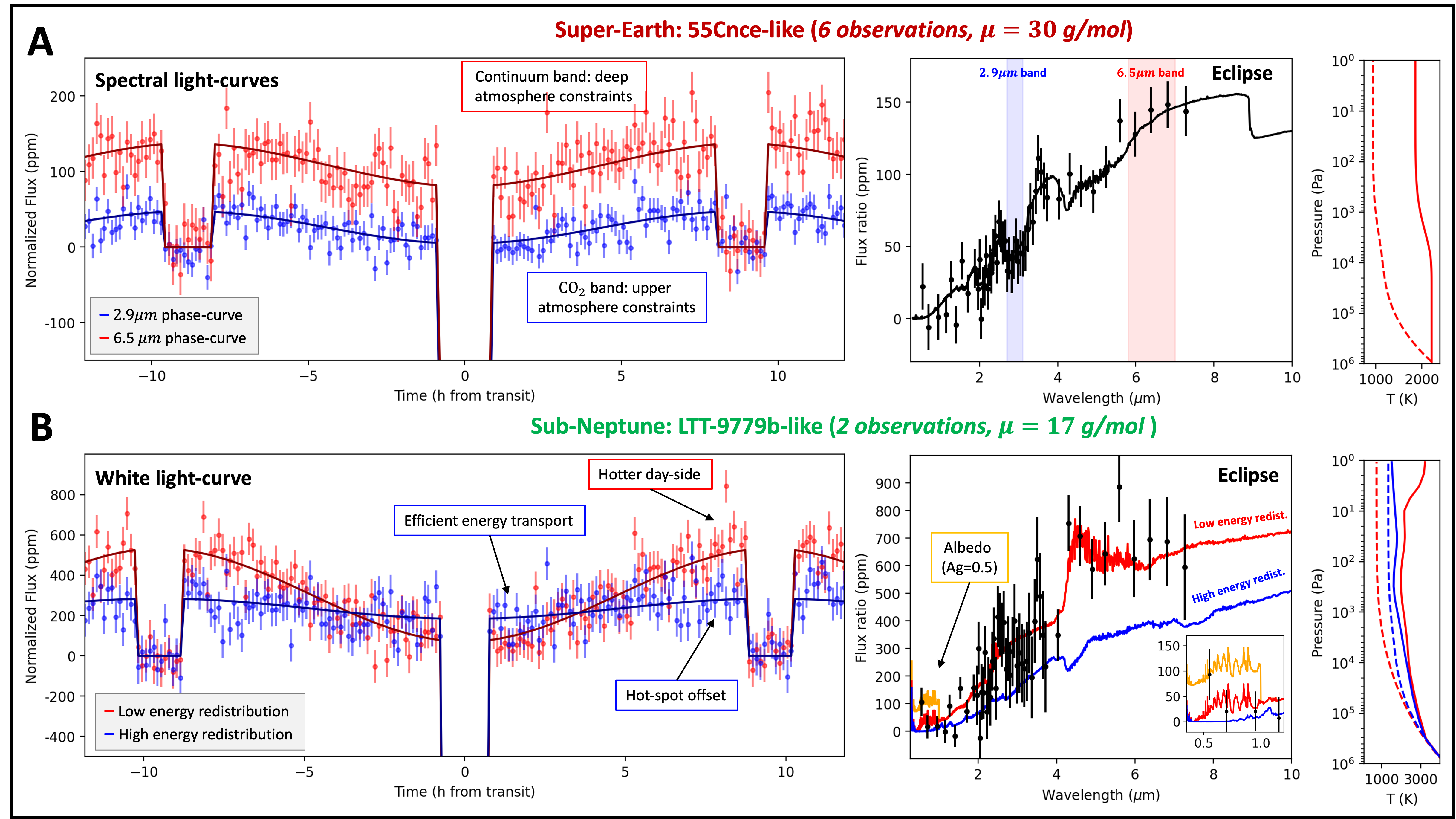}
    \caption{Simulations of phase-curves as observed by Ariel. Left panels: phase-curves; middle panels: eclipse spectra;
     right panels: thermal profiles with solid lines for the sub-stellar point and dashed line for the anti-stellar point. Panel A:   hot-rocky exoplanet  inspired from 55-Cnc\,e.  
     The planet   has a  secondary atmosphere (99\% CO + 1\% CO$_2$, from \citet{Hu_2024}) and a radius  $\sim 2.0 \, R_{\oplus}$.  Panel B: sub-Neptune  inspired from LTT-9779\,b with  a 1000x metallicity atmosphere \citep{Hoyer_2023}. We assumed $R_p \sim 4.7 \, R_{\oplus}$.}\label{fig:rockiespc}
\end{figure*}

While atmospheric studies of temperate rocky exoplanets is JWST's unique territory, as discussed in Section \ref{sec:large_collect}, studying the transition regime between sub-Neptune---possibly in the temperate regime---and warmer super-Earths is within reach of Ariel's capability. Observed features of the transition (e.g., radius valley and sub-Neptune desert) likely originate from the combination of many complex processes that cannot be uniquely constrained from mass, radius, and equilibrium temperature measurements alone. Planets in this transition regime provide a unique insight into atmospheric photochemistry, cloud/haze formation, and formation and evolutionary processes \citep{Ikoma_2006, Owen_2017, Encrenaz_2022, Kimura_2022, Lavvas_2024}. Understanding these processes require population studies of atmospheres, as a wide range of conditions needs to be tested to fully resolve correlations with e.g., host-stars' properties, systems' architecture, age. With transit spectroscopy, Ariel can efficiently constraint the atmospheric mean molecular weight for a small population of sub-Neptunes, while JWST can provide high-precision snapshots of the most interesting targets, both facilities jointly providing important clues about e.g., the origins of the radius valley and the diversity of sub-Neptunes \citep{2017AJ....154..109F, Cloutier_2020, Encrenaz_2022}. More information about the performances of Ariel's transit capabilities in this regime are provided in dedicated articles: \cite{Ito2026SmallPlanets, Venot2025ChemWG}. Other constraints could also be obtained using eclipse and phase-curve surveys. Figure \ref{fig:rockiespc} shows two examples of bright targets in this regime that could be studied through Ariel's observations: 55-Cnc\,e (panel A) and LTT-9779\,b (panel B). The figure demonstrates the advantage of phase-curves as an observing strategy for these challenging targets. Panel A (hot-rocky exoplanet case) shows that repeated spectral phase-curves would provide insight into the global structure of these atmospheres (i.e., the data are sensitive to thermal profiles and chemistry). In this example the atmosphere is assumed to be secondary, i.e., 99\% CO + 1\% CO$_2$, with $\mu = 30\,$g/mol, as inspired from \citet{Hu_2024}. Panel B (sub-Neptune case) shows how Ariel's phase-curve data could be the ideal observing mode to study the energy redistribution of small atmospheres. As discussed in Section \ref{sec:expensiveobs} and depending on the interest of the community, Ariel could be used to build a phase-curve survey a variety of close-in exoplanets, including hot super-Earths and sub-Neptunes  \cite[see also][]{Ito2026SmallPlanets}.

\subsection{Time-intensive observations: phase-curves and weather monitoring with Ariel}\label{sec:expensiveobs}

\begin{figure*}
\centering
    \includegraphics[width = 0.99\textwidth]{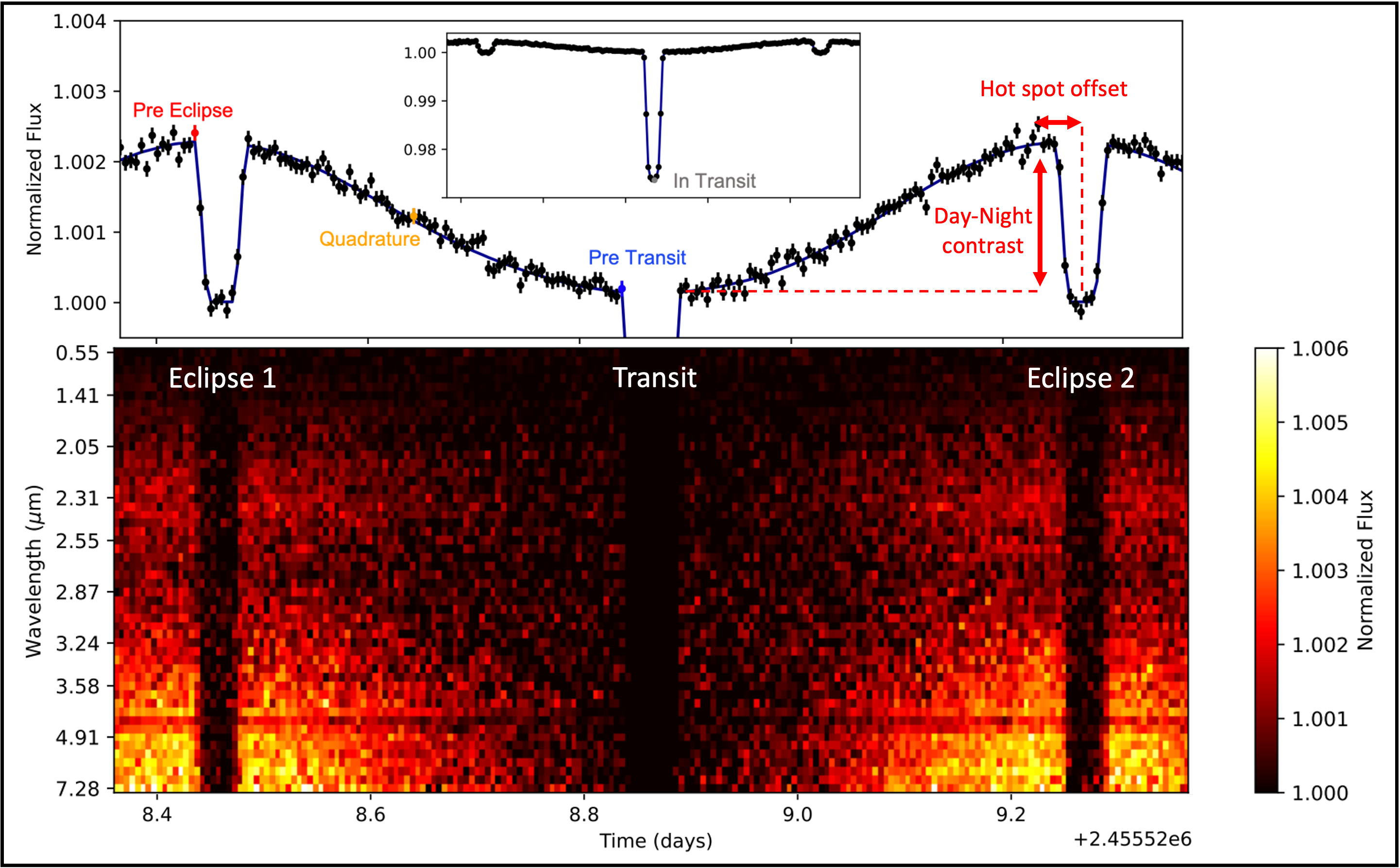}
    \caption{Simulated phase-curve of a WASP-43\,b-like exoplanet as observed with Ariel. Top: white-light curve. Bottom: spectro-temporal map of the observation. Phase-curves allow to constrain the hot-spot offset, the day-night contrast and to probe molecular absorptions (see for instance the strong CO$_2$ absorption around 4.5$\mu$m) as a function of orbital phases. Figure adapted from \citet{Changeat_2024_panchromatic} }\label{fig:wasp43pc}
\end{figure*}
\begin{figure*}
\centering
    \includegraphics[width = 0.99\textwidth]{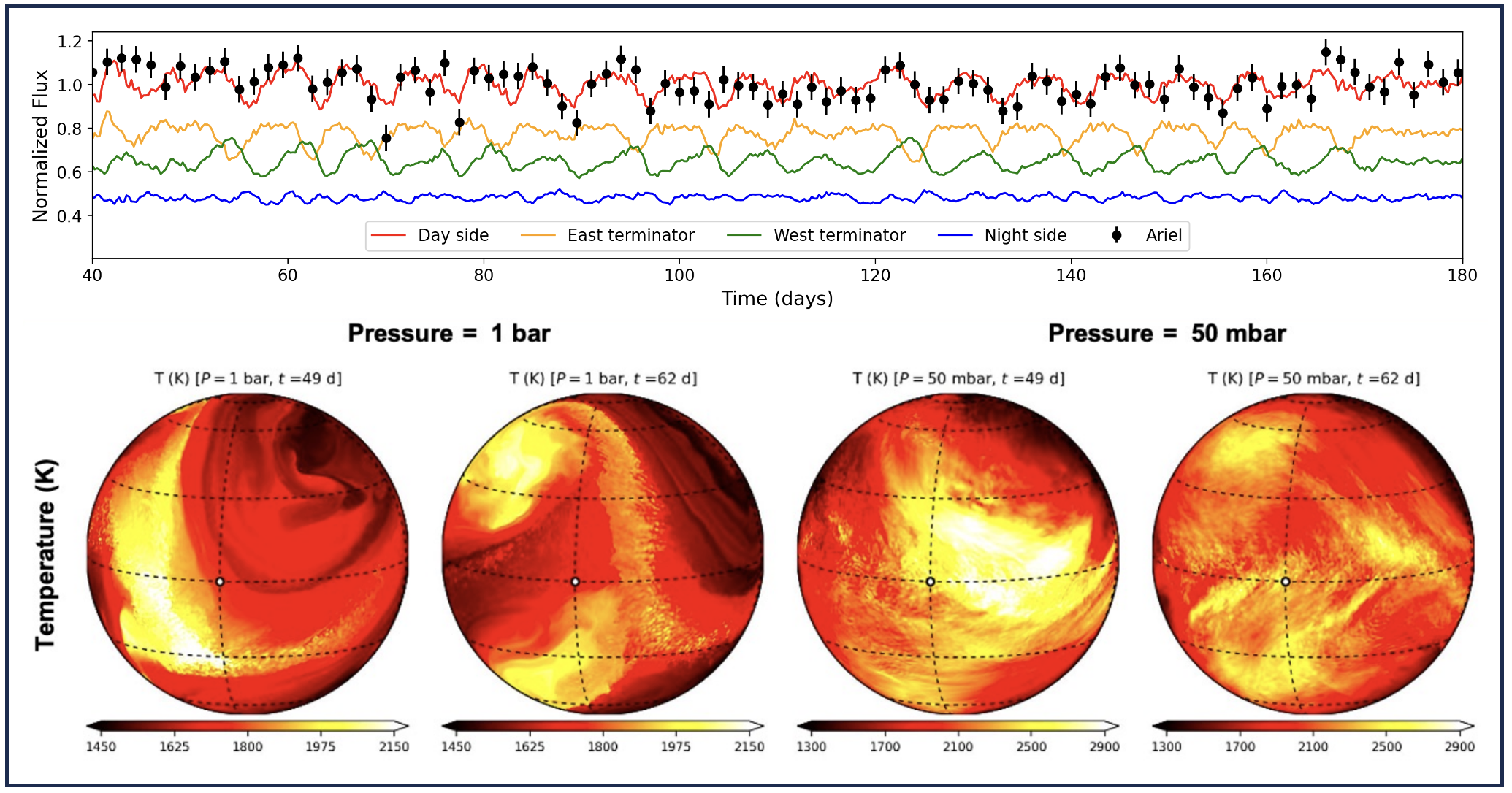}
    \caption{Blackbody emission at 10$^4$\,Pa for a ultra-hot Jupiter like WASP-121\,b (top) and example of temperature maps (bottom) from \citet{Changeat_2024}. The planetary emission is integrated from 1\,$\mu$m and 7.8$\mu$m and normalized by the mean of the day-side. Ariel's  observations and  estimated error bars are shown in black. Variability patterns of the order of a few percent  could be detected by Ariel \citep[see also ][]{Kafle_2025}.    }\label{fig:variability}
\end{figure*}

As a dedicated mission, Ariel is uniquely suited to conduct time-intensive observations  that would be difficult to schedule for general observatories like JWST. We briefly discuss below  spectroscopic phase-curves and repeated observations as examples of time-intensive observations suitable for Ariel. 
\begin{enumerate}
\item Spectroscopic phase-curves of hydrogen-rich hot exoplanets: full spectroscopic phase-curves of gaseous hot exoplanets  provide unmatched atmospheric information. Ideally two full eclipses, containing only the stellar signal, are observed, one  at the beginning and one at the end,  to anchor the entire phase-curve observation and to correct for the effects of instrument systematics. One transit event and the planetary flux as a function of orbital phases are observed in between the two eclipse events. For high SNR targets (i.e., hot Jupiters and hot Neptunes around bright stars), phase-curves provide the most complete picture of exoplanet atmospheres \citep{Teinturier_2024}. Full longitudinal constraints are important to understand the interplays between chemistry, clouds and global circulation. Hot-spot offsets or day-night contrasts (see Figure \ref{fig:wasp43pc}) could be efficiently obtained for a large population of exoplanets and irradiation conditions. As obtaining exoplanet phase-curves  requires more telescope time per target on average, these are often more difficult to schedule for a general observatory such as JWST. An optimal spectroscopic phase-curve catalogue has been identified for Ariel \cite[see e.g., the example of WASP-43\,b in Figure \ref{fig:wasp43pc} and discussions in:][]{Charnay_2022, Morales_2022}. JWST could complement said catalogue by extending the wavelength range probed, or by increasing the spectral resolution, or by expanding the population with more challenging planet-star combinations.  Additionally, faster techniques such as eclipse mapping \citep{Rauscher_2007b, Valentine_2025} could be employed---using JWST or Ariel---to provide an initial characterization before full phase-curve observations, or to build a larger catalog of phase-dependent constraints that complement phase-curve measurements. \\

\item Monitoring of atmospheric variability: exoplanets are not expected to be static objects but rather to show potentially significant weather patterns \citep{Cho_2003, Rauscher_2007, Komacek_2020}. The ability to capture such changes and to constrain their amplitude and timescales from observations is pivotal to inform current atmospheric circulation models. Also, the current practice of combining observations recorded at different epochs hinges on the assumption that atmospheric variability is not very prominent, but this might not be a realistic assumption. Studying the temporal and spatial variability of exo-atmospheres across populations is crucial to our understanding of atmospheric physical processes. Such variability can be best constrained through repeated transit, eclipse, and phase-curve observations of targets for which high SNR can be obtained with a single event (see Figure \ref{fig:variability}). Ariel Tier 3 targets are selected to optimise such investigation and provide much needed constraints 
to atmospheric variability across a broad range of parameters, including planetary temperature, stellar radiation, orbital parameters, etc. 

\end{enumerate}


\subsection{Ariel and JWST: Unique opportunities for simultaneous observations}\label{sec:simultaneous_obs}

\begin{figure*}
\centering
    \includegraphics[width = 0.8\textwidth]{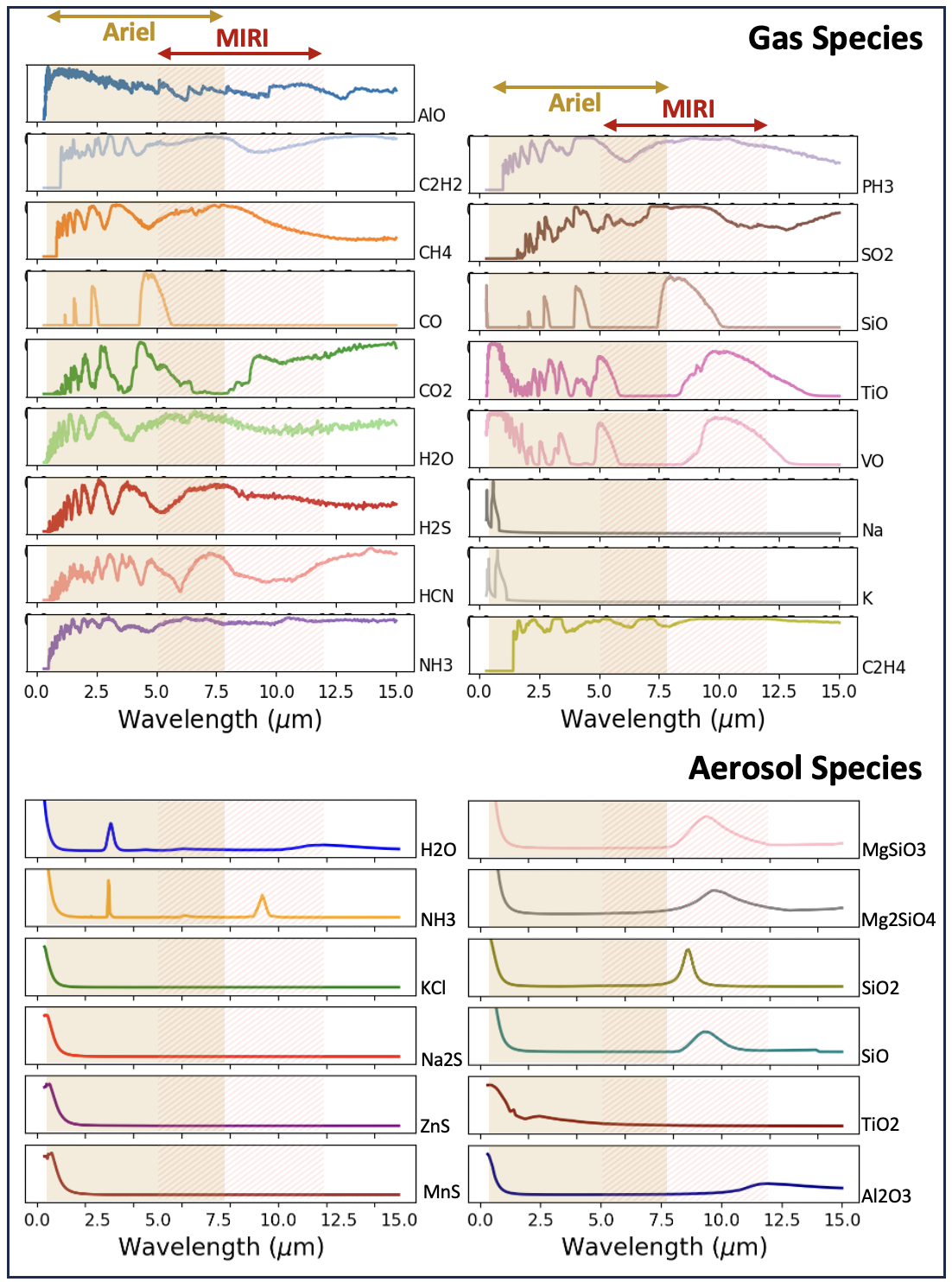}
    \caption{Normalized opacity properties of relevant aerosol species. The input aerosol optical properties are from various sources. The Mie code \textsc{PyMieScatt} \citep{Sumlin_2018} was used to compute the extinction coefficients for spherical particles of size $0.1\,\mu$m from the optical data included in \citet{batalha_2021}. See also \citet{Chubb_2024} for a summary of the Spectroscopic Database Working Group. For context, we show the wavelength regions covered by Ariel and JWST/MIRI in in shaded gold and red respectively.  }\label{fig:spectroscop}
\end{figure*}

Complementary, possibly simultaneous, observations by JWST/MIRI  could provide additional important constraints to the Ariel's  spectra. Figure \ref{fig:telescope_characteristics} and Figure \ref{fig:spectroscop} highlight the unique capabilities of JWST/MIRI-LRS to constrain, e.g., Silicate clouds around 10\,$\mu$m. Spectra covering those wavelengths are required to fully characterise  cloud/haze properties, especially if relatively large particles are present. Coordinated modeling efforts across both instruments would improve parameter robustness and reduces degeneracies. With the extended life of JWST, Ariel + JWST/MIRI-LRS observations could  be scheduled simultaneously, providing a wide 0.5--12 $\mu$m coverage and mitigating potential sources of time dependent variability caused by the star, instrument, or planetary atmosphere itself. The overlap between Ariel and MIRI-LRS  in the window 5--7.8\,$\mu$m could ease the necessary calibration between the two instruments. 

A second, compelling case for simultaneity concerns stellar contamination. The transit light source effect remains the dominant systematic for transmission spectra of planets orbiting active M~dwarfs, and it is the principal obstacle to the rocky-planet science identified as the community's top priority by \citet{Redfield_2024}. Correcting for it requires knowledge of the spot and facula covering fractions at the epoch of the observation, which is
poorly constrained by infrared spectra alone. Ariel carries continuous photometric channels down to $0.5\,\mu$m, where the contrast between the photosphere and active regions is largest. An Ariel observation scheduled simultaneously with a JWST transit/eclipse of the same target would therefore provide a contemporaneous optical constraint on the stellar surface that JWST cannot obtain for itself, at a modest cost in Ariel time. Note that this
configuration is useful even when the target is faint enough that the Ariel spectrum alone would not reach Tier2 quality.

More generally, calibration programs aiming at guaranteeing consistency in exoplanet data, could be pursued using simultaneous time-series observations with Ariel and JWST at overlapping wavelengths (i.e., with NIRISS, G395H).
Previous studies with JWST \citep{Lueber_2024} have shown, in fact, discrepancies (mainly offsets or slopes) between observations of different instruments and data reductions. 
Those differences, likely originating from the treatment of instrumental systematics (i.e., not from astrophysical origins), could be more easily investigated through joint JWST–Ariel observations. Ariel's simultaneous wavelength coverage provides a self-consistent spectral reference obtained at a single epoch, while comparison with JWST observations would enable cross-validation and help identify instrument-specific systematic effects in both facilities.

\begin{figure*}
\centering
    \includegraphics[width = 0.32\textwidth]{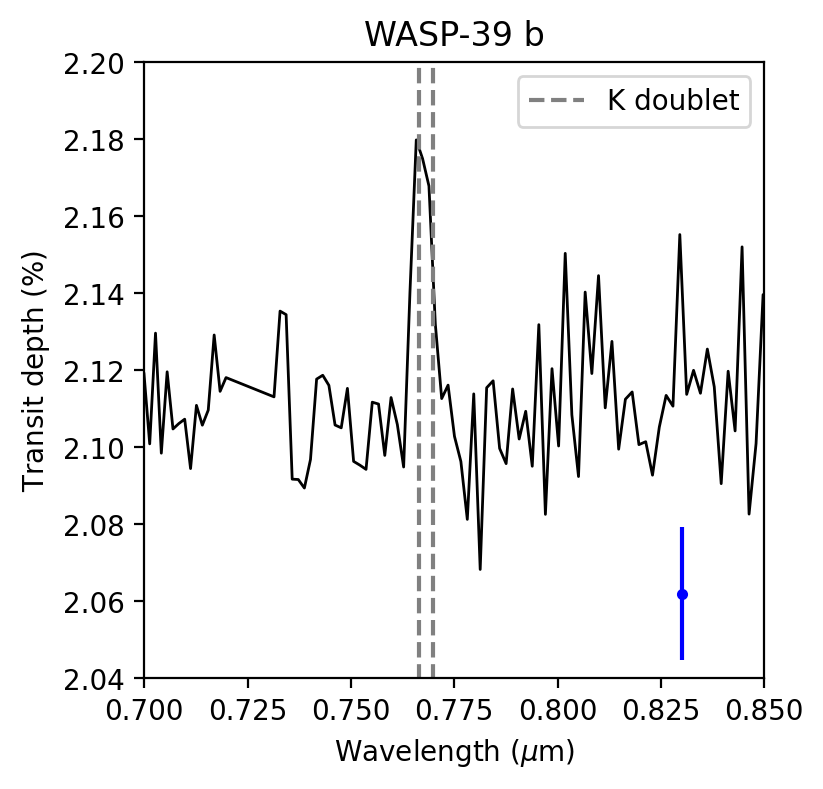}
    \includegraphics[width = 0.32\textwidth]{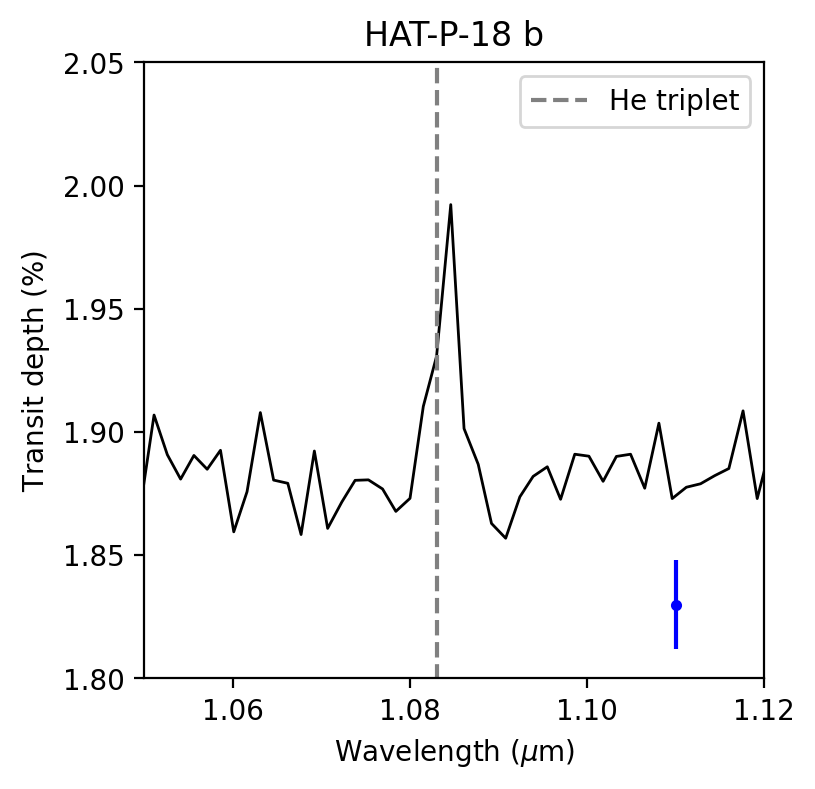}
    \includegraphics[width = 0.3\textwidth]{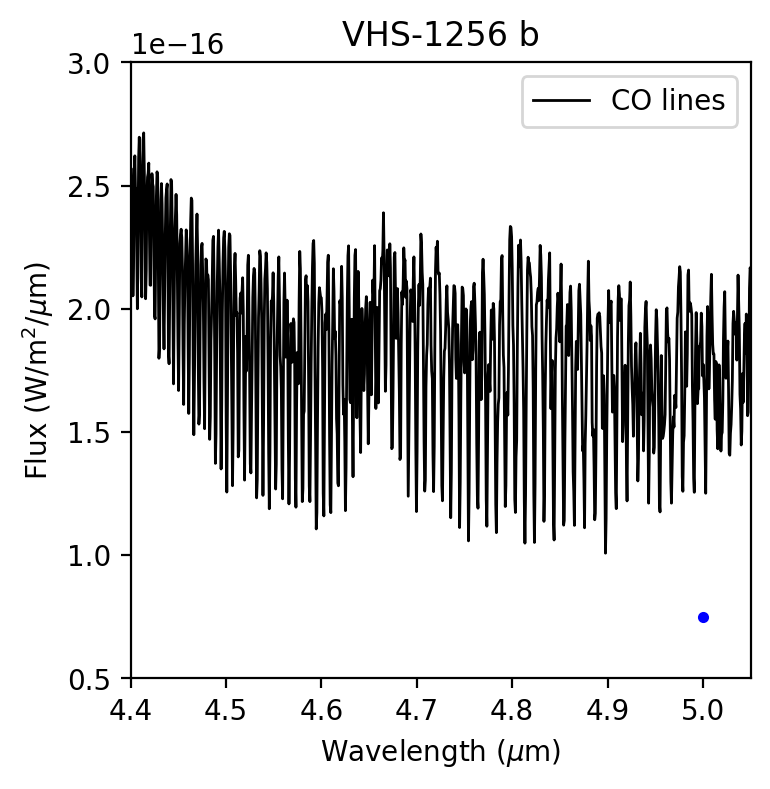}
    \caption{Examples of recent high-resolution detections of atomic and molecular species with JWST. Left: K in WASP-39\,b using NIRISS transit \citep{Holmberg_2023}. Middle: He in HAT-P-18\,b using NIRISS transit \citep{Fu_2022}. Right: CO in VHS-1256-1257\,b using NIRSpec high-contrast imaging \citep{Luhman_2023, Miles_2023}. In blue, we show representative uncertainties for this wavelength region.}\label{fig:hr_line_examples}
\end{figure*}

\subsection{JWST: an opportunity for higher spectral resolution follow-ups}\label{sec:highres}

A list of Ariel's and JWST's instruments showing their main properties (i.e., wavelength coverage and spectral resolution) is available in Figure \ref{fig:telescope_characteristics}. Ariel's spectral resolution  (i.e., R$<$150) is adequate to characterize the  continuum of exoplanet atmospheres and capture the main absorption features  of most molecular species of interest, especially at high temperature, where the spectral features are broadened. Some of JWST's instruments offer much higher spectral resolution than Ariel (up to R $\sim$ 3000),  providing   critical information about weaker transitions for an extended list of molecular species. A few examples are shown in Figure \ref{fig:hr_line_examples} and discussed below: 
\begin{enumerate}
\item Molecular characterisation. For most species, Ariel will address the issue of degeneracies through the redundancy of transitions available in its broad spectral coverage. However, the spectral resolution of JWST (R $\sim$ 3,000) is enough to resolve the lines of these species, confirm their presence and study their features in depth. Among those, there is CO (see Figure \ref{fig:hr_line_examples}), see e.g., \citep{Esparza-Borges2023,Grant2023_CO}. The characterization of CO is difficult with Ariel \citep{Changeat_2020_alfnoor} due to the relatively low resolving power of its spectrometers. Similarly, the 3.3\,$\mu$m C-H stretch can easily be observed in Ariel' spectra, probing the abundance of CH$_4$. However, its exact shape, which is also sensitive to the presence of other complex carbon chains, can be more easily constrained by JWST due to the higher resolving power. Isotopes are also within reach of JWST \citep{Barrado_2023, Gandhi_2023}. 
\item Visible light degeneracies: Ariel's photometers have sensitivity down to $\lambda = 0.5\,\mu$m, allowing to monitor stellar variability, aerosols, scattering processes, and other  absorbers in the optical (e.g., Na, K, and refractory molecules). However, the low resolution implies that some degeneracies could remain; for instance Na and K cannot be resolved by Ariel's photometers. JWST/NIRISS and JWST/NIRSpec-PRISM  provide higher spectral resolution   down to $\lambda \sim$0.6\,$\mu$m for some targets (see Figure \ref{fig:telescope_characteristics}), which could help to break these degeneracies. 
\item
 Resolving individual lines: JWST can resolve atomic or molecular lines that are not accessible to Ariel. Examples of recent detection of the K doublet in WASP-39\,b and the He triplet in HAT-P-18\,b are shown in Figure \ref{fig:hr_line_examples}. These detections provide important information on the abundance of refractory elements and atmospheric escape that are not directly accessible to Ariel's instruments.
\end{enumerate}

\section{Conclusions}

\begin{table*}
\centering
\caption{Summary of the main Ariel--JWST synergies identified in this work. For each science driver we indicate the distinctive characteristics of each observatory and some avenues for synergistic use. The last column points to the relevant section and figure of this paper.}
\label{tab:synergies}
\footnotesize
\renewcommand{\arraystretch}{1.2}
\setlength{\tabcolsep}{4pt}
\begin{tabular}{p{0.115\textwidth} p{0.195\textwidth} p{0.195\textwidth} p{0.285\textwidth} p{0.062\textwidth}}
\hline
\textbf{Science driver} & \textbf{Ariel: distinctive characteristics} & \textbf{JWST: distinctive characteristics} & \textbf{Potential synergistic use} & \textbf{Ref.} \\
\hline
\multicolumn{5}{l}{\textit{Target accessibility}} \\[2pt]
Gaseous -- bright hosts ($J \lesssim 7$) &
Full $0.5$--$7.8\,\mu$m coverage in a single visit; no saturation limit &
Detector saturation (mitigated by multistrip mode); several modes needed to match Ariel's coverage &
Ariel population for the bright targets; JWST for high-resolution follow-ups, and $\lambda > 8\,\mu$m  &
\S\ref{sec:bright}; Fig.~\ref{fig:population_ariel}\\[3pt]
Gaseous -- faint hosts ($J \gtrsim 11$) &
Requires stacking $>10$ events to reach Tier~2 quality &
NIRSpec-PRISM delivers $0.6$--$5.3\,\mu$m in a single visit &
JWST populates the faint end (combined Ariel$+$JWST sample); Ariel observes such targets only where they add sample diversity &
\S\ref{sec:bright}; Fig.~\ref{fig:population_ariel} \\[3pt]
Small, temperate planets &
Favourable planet transits can be observed, but atmospheric constraints are difficult &
Highest sensitivity for atmospeheres; fine-tuned strategies (e.g.\ MIRI $15\,\mu$m eclipses) &
JWST establishes whether atmospheres are present; Ariel places the detections within a broader population &
\S\ref{sec:large_collect},\S\ref{sec:rockies}\\[3pt]
Imaged giants \& brown dwarfs &
Not part of the survey &
High-contrast imaging and spectroscopy at very high SNR &
JWST provides constraints on young imaged giants, to compare with close-in transiting planet population from Ariel &
\S\ref{sec:large_collect} \\[3pt]
\hline
\multicolumn{5}{l}{\textit{Population-level science}} \\[2pt]
Chemical trends &
$\sim$500--1000 targets observed homogeneously over a single, fixed wavelength range &
Tens of high-precision benchmarks across heterogeneous modes &
JWST identifies candidate trends; Ariel confirms or refutes them on a statistically significant sample &
\S\ref{sec:ariel_strat},\S\ref{sec:learning}; Fig.~\ref{fig:telescope_characteristics} \\[3pt]
Super-Earth / sub-Neptune transition &
Phase-curve opportunity; small sample of transiting objects &
High-precision snapshots of individual key targets &
Joint constraints on the origin of the radius valley and of the sub-Neptune desert &
\S\ref{sec:rockies}; Fig.~\ref{fig:rockiespc} \\[3pt]
\hline
\multicolumn{5}{l}{\textit{Observing modes}} \\[2pt]
Spectroscopic phase curves &
Dedicated mission: long, repeated visits are schedulable by design &
Costly but higher SNR and resolution; extended wavelength range &
Ariel builds a phase-curve catalogue; JWST extends in wavelength, resolution, or more demanding planet--star combinations &
\S\ref{sec:expensiveobs}, Fig.~\ref{fig:wasp43pc} \\[3pt]
Temporal variability &
Tier~3 repeated visits across a broad range of planetary and stellar parameters &
Few opportunities as repeated observations are costly &
Ariel monitors variability across the population; JWST can focus on specific cases &
\S\ref{sec:expensiveobs}, Fig.~\ref{fig:variability} \\[3pt]
Eclipses \& reflected light &
Simultaneous optical--infrared coverage constrains albedo and thermal emission at one epoch &
High sensitivity to low-contrast eclipses; NIRISS reflected-light spectra; eclipse mapping &
Build a join population with wide planetary temperature and host brightness. JWST eclipse mapping to guide Ariel phase-curves &
\S\ref{sec:large_collect}, \S\ref{sec:expensiveobs} \\[3pt]
\hline
\multicolumn{5}{l}{\textit{Spectral coverage and resolution}} \\[2pt]
Clouds \& hazes &
Optical scattering slope and near-infrared continuum recorded simultaneously &
MIRI-LRS silicate features between 5 and $12\,\mu$m; NIRISS $<1\,\mu$m spectra &
Simultaneous or combined Ariel$+$MIRI-LRS observations give $0.5$--$12\,\mu$m and break cloud degeneracies (particle size, composition and abundance, temporal variability) &
\S\ref{sec:simultaneous_obs}, \S\ref{sec:highres}; Fig.~\ref{fig:telescope_characteristics},\ref{fig:spectroscop} \\[3pt]
Line-resolved chemistry &
$R < 150$: continuum, broad bands, and redundancy of transitions &
$R$ $\sim 3000$ with G395H, NIRISS: CO$_2$, CO, isotopologues, alkali, He triplet &
JWST resolves and confirms species that Ariel detects at low resolution and breaks optical degeneracies (e.g.\ Na/K) &
\S\ref{sec:highres}, Fig.~\ref{fig:hr_line_examples} \\[3pt]
Cross-calibration &
Self-consistent spectral reference obtained at a single epoch &
Multiple instruments with documented offsets &
Simultaneous observations at overlapping wavelengths to separate instrumental systematics from astrophysical variability &
\S\ref{sec:simultaneous_obs} \\[3pt]
\hline
\multicolumn{5}{l}{\textit{Methods and community}} \\[2pt]
Tools, pipelines \& models &
Inherits modeling and retrieval advancements developed for JWST; population focused tools &
High SNR drives the development of reduction, retrieval and interpretation tools &
Shared standards, benchmark datasets, open data challenges and cross-validated retrievals &
\S\ref{sec:learning} \\[3pt]
Stellar contamination &
Continuous optical photometry simultaneous with the infrared spectra &
Extensive high SNR data for active M-dwarf hosts &
Correction strategies transferred to Ariel; contemporaneous Ariel photometry can support the correction of JWST observations &
\S\ref{sec:learning}, \S\ref{sec:simultaneous_obs} \\[3pt]
Programme coordination &
Flexible target list, updated until launch and beyond &
Proposal-driven observatory; established DDT/ERS legacy &
Early coordination of target lists, joint or simultaneous programmes, shared ephemeris maintenance (ExoClock, Exoplanet Watch) &
\S\ref{sec:learning} \\
\hline
\end{tabular}\label{tab:conc}
\end{table*}

This paper summarizes recent discussions and work by contributors of the Ariel-JWST Synergy Working Group, within the Ariel Science Consortium. Building on the groundbreaking results from the first five cycles of JWST observations, we have identified here key synergies between these two observatories, which are summarized in Table \ref{tab:conc}. These synergies leverage the complementary capabilities of JWST's diverse instruments and observing modes and of Ariel's design, conceived to study efficiently a large number of exoplanet atmospheres. Despite their strong complementarity, important gaps remain that neither mission can fully address. Strategic coordination with other upcoming observatories---such as ELT, TMT, HWO, and LIFE---will be essential to advance a comprehensive understanding of exoplanetary systems.

From 2030 onward, Ariel and JWST will operate at the same time, offering new opportunities for exoplanet research. Proactive coordination is essential to fully leverage this rare observational synergy, especially given the limited and uncertain duration of their concurrent operations.
We argue that a coordinated use of JWST and Ariel, either through joint---possibly simultaneous---programs or by prioritizing  observations which make use of their unique performances, would benefit the entire exoplanet field and enable even more groundbreaking discoveries.

\section*{Acknowledgements}

We thank the two anonymous reviewers for their constructive comments and suggestions, which helped improve this manuscript. This publication received financial support from CNES (PI: P-O. Lagage). This publication is part of the project ``Exoplanet Atmospheres with Next-generation Space Telescopes'' with file no. VI.Veni.242.091 (PI: Q. Changeat) of the ``NWO Talent Programme Veni Science domain 2024'' under the grant \url{https://doi.org/10.61686/QPZSS86131}. It is also part of the project ``Interpreting exoplanet atmospheres with JWST'' with file no. 2024.034 (PI: Q. Changeat) of the research programme ``Rekentijd nationale computersystemen'' which is (partly) financed by the Dutch Research Council (NWO) under the grant \url{https://doi.org/10.61686/QXVQT85756}. This work used the Dutch national e-infrastructure with the support of the SURF Cooperative using grant no. 2024.034. It has also received funding from the ESA-OSIP program, contract no. 4000152963 (PI: Q. Changeat). We also acknowledge the availability and support from the High-Performance Computing platforms (HPC) from DIRAC, and OzSTAR, which provided the computing resources necessary to perform this work. This work utilised the Cambridge Service for Data-Driven Discovery (CSD3), part of which is operated by the University of Cambridge Research Computing on behalf of the STFC DiRAC HPC Facility (\url{www.dirac.ac.uk}). The DiRAC component of CSD3 was funded by BEIS capital funding via STFC capital grants ST/P002307/1 and ST/R002452/1 and STFC operations grant ST/R00689X/1. DiRAC is part of the National e-Infrastructure. This work utilised the OzSTAR national facility at Swinburne University of Technology. The OzSTAR program receives funding in part from the Astronomy National Collaborative Research Infrastructure Strategy (NCRIS) allocation provided by the Australian Government.
C.D. acknowledges financial support from the grant RYC2023-044903-I funded by MCIU/AEI/10.13039/501100011033, the ESF+, and from the INAF initiative ``IAF Astronomy Fellowships in Italy'', grant name \textit{GExoLife}.
G.M. acknowledges financial support from the Severo Ochoa grant CEX2021-001131-S and from the Ramón y Cajal grant RYC2022-037854-I funded by MCIN/AEI/1144 10.13039/501100011033 and FSE+. A.B. was supported by the Italian Space Agency (ASI) with Ariel grant n. 2021.5.HH.0.
O.V. acknowledges funding from the ANR project `EXACT' (ANR-21-CE49-0008-01) and from the Centre National d'Etudes Spatiales (CNES).
PL was supported by the Action Thématique Ex- 860 osystèmes of CNRS/INSU, co-funded by CEA and CNES.

\section*{Data Availability}

The data underlying this article will be shared on reasonable request to the corresponding author. \\

\noindent Authors declare no conflict of interest



\bibliographystyle{rasti}
\bibliography{main} 




\appendix

%
%
%

\clearpage

\appendix

\section{Atmospheric models for the simulated spectra of WASP-39\,b and HD-189733\,b}
\label{app:transit}

The Ariel transmission spectra shown in the bottom left panels of Figure\,\ref{fig:telescope_characteristics} are forward models computed with the open-source atmospheric framework \textsc{TauREx}\,3.1 \citep{2019_al-refaie_taurex3, al-refaie_2021_taurex3.1}. In both cases the atmosphere is described in plane-parallel transmission geometry, discretised over 100 layers that are equally spaced in $\log p$ between $p=10^{6}$~Pa and $p=10^{-2}$~Pa. The forward model is first evaluated at the native resolution of the input cross-sections, i.e., $\mathcal{R} = 15,000$, and only binned down at the last step (see Section\,\ref{app:transit_noise}).

\subsection{Opacity sources}
\label{app:transit_opacities}

In the forward model, the following extinction sources are included: molecular absorption, collision
induced absorption (CIA) from the H$_2$--H$_2$ and H$_2$--He pairs \citep{abel_h2-h2, abel_h2-he, fletcher_h2-h2}, Rayleigh scattering \citep{cox_allen_rayleigh}, and Mie extinction by cloud particles (see Section\,\ref{app:transit_clouds}). Molecular cross-sections are pre-sampled at a resolution of
$\mathcal{R}=15\,000$ over $\lambda\in[0.3,50]\,\mu$m from the \textsc{ExoMol} project line lists \citep{tennyson2012exomol, Tennyson_exomol, Chubb_2021_exomol}; the species entering the
calculation are: H$_2$O, CO, CO$_2$, CH$_4$, SO$_2$, TiO and VO.

\subsection{System parameters, thermal structure and chemistry}
\label{app:transit_chemistry}

\underline{WASP-39\,b}: For the faint target we adopt the bulk parameters of WASP-39\,b
($R_{\rm p}=1.27\,R_{\rm J}$, $M_{\rm p}=0.28\,M_{\rm J}$) orbiting a star of radius $R_\star=0.895\,R_\odot$ \citep{Bonomo_2017}. The thermal structure is assumed isothermal at $T=1022$\,K, consistent with the equilibrium temperature of the planet and atmospheric retrieval results. The chemical composition is computed at thermochemical equilibrium using the \textsc{GGchem} code \citep{Woitke_2018_GG}, interfaced through the \textsc{taurex-ggchem} plugin \citep{al-refaie_2021_taurex3.1}, assuming a metallicity of 20 times solar and solar C/O ratio. On top of the equilibrium solution we impose a constant-with-altitude volume mixing ratio of 10\,ppm for SO$_2$, in order to mimic the photochemical production of this species now routinely inferred from JWST observations of warm gaseous planets \citep{Rustamkulov_2023, Dyrek_2023, Powell_2024}: SO$_2$ is a disequilibrium product and is therefore not captured by the equilibrium calculation. All other abundances---including
those of TiO and VO, which shape the optical continuum---follow directly from the equilibrium computation. 

\noindent\underline{HD-189733\,b}: For the bright target we adopt the bulk parameters of HD-189733\,b
($R_{\rm p}=1.13\,R_{\rm J}$, $M_{\rm p}=1.13\,M_{\rm J}$) orbiting a star of radius $R_\star=0.75\,R_\odot$, with an isothermal profile at $T=1200$\,K \citep{Stassun_planetparam}. The chemistry is again computed with \textsc{GGchem}, here at solar metallicity and C/O ratio. Since this planet is cooler and less metal-rich than the WASP-39\,b like case, we do not include SO$_2$ (the species is also not detected in JWST observations). \\

Note that these models are used to illustrate our discussion of the JWST/Ariel synergies, but they are not intended to perfectly reproduce available JWST observations nor to be used for direct interpretations of real spectra. 

\subsection{Clouds}
\label{app:transit_clouds}

Parameterized aerosols are added to the models of WASP-39\,b and HD-189733\,b using the \textsc{taurex-PCQ} plugin \citep{Changeat_2025, Voyer_2024}, which computes the extinction efficiency $Q_{\rm ext}$ of spherical particles from tabulated complex refractive indices using \textsc{PyMieScatt} \citep{Sumlin_2018}. In each case we include a single aerosol layer of spherical particles, with optical constants taken from the \textsc{Virga} compilation of \citet{batalha_2021}: amorphous MgSiO$_3$ for the WASP-39\,b like case and SiO$_2$ for the HD-189733\,b like case. Both are plausible condensates in this temperature regime and both have recently been inferred from JWST observations of hot gaseous planets \citep{Grant_2023, Dyrek_2023}. The particle size distribution follows the one-parameter form of \citet{Budaj_2015}, uniquely defined by a critical radius $\mu_r$ and sampled at five radii when integrating Equation\,(A.4) of \citet{Changeat_2025}. Each layer is centred at a pressure $P$ and spans one pressure decade, with a particle number density that
decays exponentially with pressure from a reference value $\chi$ at the base
of the layer. The adopted values are selected to roughly agree with current JWST observations of these objects and are listed in Table\,\ref{tab:transit_params} for reproducibility.

\subsection{Instrument simulation}
\label{app:transit_noise}

Ariel uncertainties are estimated with the official radiometric model
\textsc{ArielRad} \citep{Mugnai_2020}, using the following versions: \textsc{ArielRad} v2.4.26, \textsc{ExoRad}
v2.1.111, Payload v0.0.17. The high-resolution model is
binned onto Ariel Tier3 resolution and the per-channel uncertainty is added. In figure \ref{fig:telescope_characteristics}, we show a single Gaussian realisation of the simulated Ariel spectra. The underlying solid black line is the same forward model binned to $\mathcal{R}=500$ for display. Finally, we stress that the JWST datapoints overlaid in Figure~\ref{fig:telescope_characteristics} are from real observations \citep{Carter_2024, Powell_2024, Zhang_2025} and not simulations.

\begin{table}
\caption{Parameters of the \textsc{TauREx} forward models used to produce the
simulated Ariel transmission spectra of Figure~\ref{fig:telescope_characteristics}.}
\label{tab:transit_params}
\begin{tabular}{lcc}
\hline
Parameter & WASP-39\,b like & HD~189733\,b like \\
\hline
$R_{\rm p}$ [$R_{\rm J}$]              & 1.27  & 1.13 \\
$M_{\rm p}$ [$M_{\rm J}$]              & 0.28  & 1.13 \\
$R_{\star}$ [$R_{\odot}$]              & 0.895 & 0.75 \\
$T$ [K] (isothermal)                   & 1022 & 1200 \\
Metallicity                            & $20\times$ solar & solar \\
C/O                                    & solar & solar \\
$\log$(SO$_2$) (imposed)               & $-5$ & None \\
Cloud species                          & MgSiO$_3$ (am.) & SiO$_2$ \\
$\mu_r$ [$\mu$m]                       & 0.5 & 0.31 \\
$P_{\rm cloud}$ [Pa]                   & $10$ & $10^{2}$ \\
$\Delta P_{\rm cloud}$ [dex]           & 1 & 1 \\
$\chi$ [m$^{-3}$]                      & $10^{5}$ & $2\times10^{5}$ \\
$N_{\rm layers}$                       & 100 & 100 \\
$N_{\rm transits}$                     & 2 & 6 \\ 
\hline
\end{tabular}
\end{table}

\section{Atmospheric models for the simulated Ariel phase-curves of 55-Cnc\,e and LTT-9779\,b}
\label{app:phasecurve}

\subsection{The 1.5D phase-curve model}
\label{app:pc_model}

The phase-curves of Figure\,\ref{fig:rockiespc} are computed with the 1.5D
phase-curve model of \citet{changeat_2021_phasecurve2}, as implemented in the \textsc{ExPLOR} code \citep{Changeat_2024_panchromatic}, which also
inherits from the \textsc{TauREx}\,3.1 framework.
The model represents the planet as three regions of homogeneous properties---a hot-spot, a dayside and a nightside---each with its own thermal structure,
chemical profiles and set of opacity contributions, which may be set
independently or coupled. For this article, we combine hot-spot and day-side regions, modeling planets with only two regions that split at the terminator. At each orbital phase the emission of the visible hemisphere is obtained using a Gaussian quadrature scheme for the radiative transfer, here using $N_{\rm gauss}=40$ nodes; we refer the reader to \citet{Changeat_2020_phasecurve1}, \citet{changeat_2021_phasecurve2} for the detailed formulation.

In Figure\,\ref{fig:rockiespc} the radiative transfer is performed at native cross-section resolution and the orbit is sampled with 200 epochs of equal duration ($\sim 7$\,min) covering one complete orbital revolution plus a baseline of a few transit durations on either side of the transit. This sampling therefore contains two eclipses, the transit, and the out-of-event baselines required to anchor the observation, which is the baseline phase-curve strategy foreseen for Ariel \citep{Tinetti_2021_redbook} and the strategy adopted by most JWST programmes.

\subsection{Atmospheric model for 55-Cnc\,e}
\label{app:pc_55cnce}

The 55-Cnc\,e-like simulation is inspired from recent results of \citet{Hu_2024}. We adopt the system parameters from \cite{bourrier}: $R_{\rm p}=1.9\,R_{\oplus}$ and $M_{\rm p}=8.0\,M_{\oplus}$, $T_\star=5172$\,K, and $R_\star=0.94\,R_\odot$.

The atmosphere is assumed to be secondary and CO-dominated, with a background mixture of CO and CO$_2$ at a volume mixing ratio of 1\%, giving a mean molecular weight of $\mu\simeq30$~g\,mol$^{-1}$. This composition is inspired by the secondary atmosphere inferred by \citet{Hu_2024}. The thermal structure of the dayside hemisphere is described by the analytic profile of \citet{Guillot_TP_model} with $T_{\rm irr}=2000$\,K, while the nightside uses $T_{\rm irr}=1000$\,K and $T_{\rm int}=500$~K, producing the day--night contrast visible in the right-hand panel of Figure\,\ref{fig:rockiespc}. The opacity sources are molecular absorption (CO and CO$_2$), Rayleigh
scattering, and CO$_2$--CO$_2$ CIA.

In Figure\,\ref{fig:rockiespc}, the spectral light-curves shown in the left panel are obtained by integrating the spectro-temporal map over two synthetic photometric bands, $\lambda\in[2.7,3.1]\,\mu$m and $\lambda\in[5.8,7.0]\,\mu$m, chosen to maximize the contrast between a continuum region---which probes the deep atmosphere---and the CO$_2$ band, which probes higher altitudes.

\subsection{Atmospheric model of LTT-9779\,b}
\label{app:pc_ltt}

The LTT-9779\,b-like simulations are inspired from recent results of \citet{Hoyer_2023}. We adopt the system parameters from \cite{Jenkins_2020}: $R_{\rm p}=4.7\,R_{\oplus}$, $M_{\rm p}=29\,M_{\rm \oplus}$, $T_\star=5480$\,K, and $R_\star=0.95\,R_\odot$. 

The chemistry is computed at thermochemical equilibrium with \textsc{GGchem} code \citep{Woitke_2018_GG}, assuming a metallicity of 1000 times solar as suggested by \citet{Hoyer_2023}; this yields a mean molecular weight of $\mu\simeq17$\,g\,mol$^{-1}$. Opacities include molecular absorption, Rayleigh scattering and H$_2$--H$_2$ and H$_2$--He CIA.

We build two scenarios, differing only in their energy redistribution.
For each scenario, the day-side thermal structure is taken directly from the self-consistent profiles of \citet{Hoyer_2023}: the no-redistribution profile for the \emph{low redistribution} case, and the $f=1.5$ profile for the \emph{high redistribution} case. For the nightside, no profiles are available, so we choose to model them using \citet{Guillot_TP_model} and construct arbitrary case to illustrate our main points. In the low redistribution case, the nightside receives no stellar energy, so we use $T_{\rm irr}=0$\,K and $T_{\rm int}=870$\,K. The temperature $T_{\rm int}$ is chosen to match day and night side profiles at 3\,bar \cite[the deepest pressure in][]{Hoyer_2023}. In the high redistribution case we use $T_{\rm irr}=1400$\,K to mark the transport of energy from day to night side, and $T_{\rm int}=870$\,K.

The \textsc{TauREx} emission model does not include starlight scattered by the planetary atmosphere, which for a highly reflective planet such as LTT-9779\,b is a significant contribution at optical wavelengths \citep{Hoyer_2023}. We therefore add a posteriori a grey reflected component to
the eclipse spectrum,
\begin{equation}
\left(\frac{F_{\rm ref}}{F_{\star}}\right)(\lambda) = A_{\rm g}(\lambda)
\left(\frac{R_{\rm p}}{a}\right)^{2},
\end{equation}
with a geometric albedo $A_{\rm g}=0.5$ for $\lambda<1\,\mu$m and $A_{\rm g}=0$ at longer wavelengths. This component is the origin of the optical excess highlighted in the inset of Figure~\ref{fig:rockiespc}, and illustrates the diagnostic value of Ariel's simultaneous optical coverage for this class of object.

The white light-curve shown in the left panel is obtained by integrating the spectro-temporal map over the full $\lambda\in[2.0,7.8]\,\mu$m range.

\subsection{Instrument simulation}
\label{app:pc_noise}

As for the transit simulations, Ariel uncertainties are estimated with \textsc{ArielRad} \citep{Mugnai_2020}. The per-channel uncertainties are propagated through the same binning as the model when constructing the broad-band light-curves, and scaled by
the number of phase-curve observations, here six for 55-Cnc\,e and two for LTT-9779\,b. The
plotted datapoints are a single Gaussian realisation of the model with those uncertainties. Because each phase-curve contains two eclipses, the eclipse spectra shown in the middle panels are stacked over a correspondingly larger number of events.

\begin{table}
\caption{Parameters of the \textsc{ExPLOR} phase-curve models used to produce Figure~\ref{fig:rockiespc}. HS: hot-spot; D: dayside;
N: nightside.}
\label{tab:pc_params}
\begin{tabular}{lcc}
\hline
Parameter & 55~Cnc\,e like & LTT-9779\,b like \\
\hline
$R_{\rm p}$ [$R_{\oplus}$]        & 1.9 & 4.7 \\
$M_{\rm p}$ [$M_{\oplus}$]         & 8.0 & 29 \\
$R_{\star}$ [$R_{\odot}$]        & 0.94 & 0.95 \\
$T_{\star}$ [K]                   & 5172 & 5480 \\
Chemistry                         & CO/CO$_2$ (fixed) & $Z=1000\times$ solar \\
$\mu$ [g\,mol$^{-1}$]             & 30 & 17 \\
$T$--$p$ (day)                  & Guillot, $T_{\rm irr}=2000$~K & From \citet{Hoyer_2023} \\
                                & $T_{\rm int}=100$~K & \\
$T$--$p$ (night)                      & Guillot, $T_{\rm irr}=1000$~K, & Guillot\footnotemark \\
                                  & $T_{\rm int}=500$~K            &  \\
$N_{\rm layers}$                  & 100 & 100 \\
$p$ range [Pa]                    & $10^{-2}$--$10^{6}$ & $1$--$10^{6}$ \\
$N_{\rm gauss}$                   & 40 & 40 \\
$N_{\rm phases}$                  & 200 & 200 \\
$N_{\rm obs}$                     & 6 & 2 \\
\hline
\end{tabular}
\end{table}
\footnotetext{For low redistribution case: $T_{\rm int}=870$\,K, $T_{\rm irr}=0$\,K; For high redistribution case: $T_{\rm int}=870$\,K, $T_{\rm irr}=1400$\,K.}



\bsp	
\label{lastpage}
\end{document}